\documentclass[
authoryear, 12pt]{elsarticle}

\usepackage[a4paper]{geometry}
\usepackage{amssymb}

\usepackage{changepage}
\usepackage{setspace}

\usepackage{chngcntr}
\usepackage{caption}
\usepackage{psfrag}
\usepackage{graphicx}
\usepackage{latexsym}
\usepackage{keyval}
\usepackage{ifthen}
\usepackage{moreverb}
\usepackage{gnuplottex}
\usepackage{upgreek}
\usepackage{subfig}
 \usepackage{amsmath}
 \usepackage{mathtools}
\usepackage{appendix}
\usepackage{color}
\usepackage[usenames,dvipsnames]{xcolor}
\usepackage{enumerate}
\usepackage{wrapfig}
\usepackage{amssymb}
\usepackage{natbib}
\usepackage{algorithm}
\usepackage{algpseudocode}
\usepackage[hidelinks]{hyperref}

\usepackage{titlesec}
\titlelabel{\thetitle \quad}

\usepackage{ulem}
\newcommand{\bx}{\bar \fx}
\newcommand{\fzero}{\sty{ 0}}
\newcommand{\bgrad}[1]{\overline{\rm grad}\left( #1 \right)}
\newcommand{\Nph}{N_{\rm ph}}
\newcommand{\avg}[1]{\langle#1\rangle}
\newcommand{\jmp}[1]{[\![#1]\!]}
\newcommand{\fEps}{\mbox{\boldmath $\mathcal{E} $}}
\newcommand{\tcEps}{\tilde {\sty \fEps}}
\newcommand{\bcEps}{\bar {\sty \fEps}}
\newcommand{\NipFE}{N_{\rm ip}}

\newcommand{\Nmd}{N_{\rm md}}
\newcommand{\Nmdsgma}{N_{\rm md}^\sigma}
\newcommand{\Nsim}{N_{\rm sim}}

\newcommand{\infi}[1]{\underset{{#1}}{\rm inf}}

\newcommand{\fepsp}{\feps^{\rm p}}
\newcommand{\fepse}{\feps^{\rm e}}
\newcommand{\sigmaD}{\sigma^{\rm D}}
\newcommand{\sigmavM}{\sigma_{\rm vM}}
\newcommand{\sigmay}{\sigma_{\rm y}}
\newcommand{\barCa}{\bar\ffC^{\rm a}}
\newcommand{\Ca}{\ffC^{\rm a}}

\newcommand{\Ds}{\displaystyle}

\newcommand{\blds}[1]{\mbox{\scriptsize \boldmath $#1$}}

\newcommand{\T}[1]{{#1}^{\sf  T}}

\newcommand{\pd}[2]{\displaystyle\frac{\partial #1}{\partial #2}}

\newcommand{\normg}[1]{\left\Vert \, #1 \, \right\Vert}

\newcommand{\grad}[1]{{\rm grad}\left( #1 \right)}
\renewcommand{\div}[1]{{\rm div }\left( #1 \right)}

\newcommand{\fsym}[1]{{\rm sym }( #1 )}

\newcommand{\unit}[1]{\,{\rm #1}}

\newcommand{\cblue }{\color{blue}}

\newcommand{\fempty}[1]{{}}

\newcommand{\f}[1]{\mbox{$ #1 $}}

\newcommand{\sty}[1]{\mbox{\boldmath $#1$}}
\newcommand{\styy}[1]{{\mathbb{#1}}}

\newcommand{\fn}{\sty{ n}}

\newcommand{\ft}{\sty{ t}}
\newcommand{\fu}{\sty{ u}}

\newcommand{\fx}{\sty{ x}}

\newcommand{\fA}{\sty{ A}}

\newcommand{\fD}{\sty{ D}}
\newcommand{\fE}{\sty{ E}}

\newcommand{\fH}{\sty{ H}}
\newcommand{\fI}{\sty{ I}}

\newcommand{\fN}{\sty{ N}}

\newcommand{\fR}{\sty{ R}}
\newcommand{\fS}{\sty{ S}}

\newcommand{\fX}{\sty{ X}}

\newcommand{\ffC}{\styy{ C}}

\newcommand{\ffI}{\styy{ I}}

\newcommand{\ffP}{\styy{ P}}

\newcommand{\ffR}{\styy{ R}}

\newcommand{\fsigma}{\mbox{\boldmath $\sigma$}}

\newcommand{\fxi}{\mbox{\boldmath $\xi $}}
\newcommand{\feta}{\mbox{\boldmath $\eta $}}

\newcommand{\feps}{\mbox{\boldmath $\varepsilon $}}

\newcommand{\cB}{{\cal B}}

\newcommand{\tcfE}{\tilde{\sty{{\cal E}}}}

\newcommand{\utcE}{\underline{\tilde{{\cal E}}}}

\renewcommand{\d}{{\,\rm  d}}
\newcommand{\bdiv}[1]{\overline{\rm div }\left( #1 \right)}

\begin{document}
\pagenumbering{arabic}

\newboolean{badQuality}
 \setboolean{badQuality}{false}

\begin{frontmatter}





\title{Sequential Subspace Mode Adaptation for the Reduced-Order Homogenization of Dissipative Microstructures using E3C Hyper-Reduction}

\author[1]{Hauke Goldbeck}
\author[1]{Stephan Wulfinghoff}

\address[1]{Computational Materials Science, Department of Materials Science, Kiel University, Kaiserstr.~2, 24143 Kiel, Germany\\
\{hago,swu\}@tf.uni-kiel.de}


\begin{abstract}
Three-dimensional inelastic computational homogenization of complex engineering components requires a multitude of nonlinear microstructural simulations, making it computationally expensive. This work investigates a projection-based model order reduction (pMOR) method with 'Sequential Subspace Mode Adaptation', which can be easily integrated into existing codes using linear subspaces. Starting with a 'conventional' linear subspace strain approximation, the dynamic online construction of a second -- lower dimensional -- affine subspace embedded in the linear subspace determined offline leads to a further reduction of the dimensionality.\\
A second novelty is the outline of the E3C hyper-reduction method for non-crystalline dissipative materials with internal variables, introducing a viscous regularization of non-differentiable stress-strain relations. In addition, a theoretical discussion is provided, illustrating that the E3C method aims at satisfaction of a projected and hyper-reduced variant of the classical Hill-Mandel macro-homogeneity condition. The latter theoretically implies equivalence with the high-dimensional model and satisfaction of both the hyper-reduced weak equilibrium and compatibility conditions.\\
The influence of training batch size, material nonlinearity, and microstructure on the performance are evaluated through parameter studies. Three-dimensional elastoplastic two-scale simulations with hundreds of thousands of macroscopic degrees of freedom illustrate the efficiency and accuracy, with computational times approaching those of single scale simulations.
\end{abstract}

\begin{keyword}
Hyper-reduction \sep Reduced order models \sep Computational homogenization \sep E3C \sep Sequential subspace mode adaptation
\end{keyword}

\end{frontmatter}

\section{Introduction}
In recent decades, inelastic simulations of engineering structures have become an industrial standard. The main ingredients of these simulations are {\it i)} the finite element method (FEM), which is used to solve {\it ii)} the linear momentum balance and {\it iii)} a set of constitutive equations. While the validity of the first two ({\it i} and {\it ii}) is widely accepted, the constitutive equations are often considered the main origin of deviations from experimental results. Therefore, instead of assuming explicit inelastic constitutive laws with limited accuracy, recent research focuses on more flexible frameworks, like the automated identification of the constitutive model, selecting a small number of model terms from a large number of candidate expressions \citep{flaschel2023automated}, or the use of physics-augmented artificial neural networks for viscoelasticity \citep{kalina2025physics}, elastoplasticity \citep{boes2025accounting}, general inelastic constitutive artificial neural networks \citep[ICANNs, see][]{holthusen2024theory, holthusen2026complement} or convex neural networks approximating the potentials of generalized standard materials \citep{flaschel2025convex}. The training data required to calibrate the aforementioned models often stems from experimental data or artificial mechanical tests using a microscopic computational model, resolving the material's microstructure, in the sense of a 'top-down' approach.\\
An alternative to macroscopic material modeling is computational homogenization, i.e., the direct coupling of the microscopic and macroscopic models within a two-scale approach, and thus predicting 'bottom up' the macroscopic mechanical response, using appropriate scale transition techniques. Classically associated with the FE\f{^2}-method \citep{feyel1999multiscale,miehe1999computational}, computational homogenization methods are computationally comparably expensive due to the large number of microscopic problems which need to be solved. Model order reduction (MOR) techniques decrease the dimensionality of the microscale model, and consequentially the computational effort, while preserving essential microscopic physical features like the equilibrium conditions or the material laws of the microscopic constituents. There are also hybrid modeling approaches, incorporating features of both, MOR and neural networks. For example, Deep Material Networks \citep{liu2019deep} exhibit a layered structure, similar to that of artificial neural networks. Mimicking equilibrium states of hierarchical laminates, their building blocks are the microstructural material laws. Once trained on a certain microstructure, the constitutive equations of the phases may be exchanged, preserving a remarkable accuracy without retraining \citep{srinivas2026rapid}.\\
Reduced order models (ROMs) require training data -- the so-called 'snapshots', i.e., simulation results obtained from a high-dimensional microscopic model (e.g., an FEM- or FFT-model), based on which a compressed model is derived. This can be achieved by clustering techniques \citep{cavaliere2020efficient,ferreira2025finite} or -- more frequently -- using the proper orthogonal decomposition \citep[POD, e.g.,][]{holmes1996turbulence}. Related methods extract global shape functions (the 'modes') from the snapshot data, usually by singular value decomposition (SVD), and thus compress the high-dimensional degrees of freedom into a low-dimensional subspace, being parametrized by the mode coefficients. An early contribution in the homogenization context is the nonuniform transformation field analysis (NTFA) by \citet{michel2003nonuniform}, which generalizes the TFA by \citet{dvorak1992transformation}, allowing for nonuniform modes of the internal variables and their driving forces \citep{fritzen2016finite}. Further reduction of the computational effort is achieved \citep[e.g.][]{michel2016model,michel2017effective}, exploiting ideas that find their origin in nonlinear homogenization theory \citep[e.g.][]{castaneda1996exact,ponte2015fully}.\\
Alternatively, global displacement or displacement fluctuation modes \citep{yvonnet2007reduced} are also heavily used in combination with a conventional Galerkin projection. In other words, the microscopic equilibrium condition is projected onto the lower-dimensional linear subspace spanned by the displacement fluctuation modes, hence the name 'projection-based ROMs' (pROMs). This projection step requires the numerical evaluation of integrals over the whole microstructure, and becomes the computational bottleneck, if the nonlinear material law is evaluated at all integration points of the high-dimensional model (sometimes also called 'high-fidelity model'). Hyper-reduced order models (hROMs) aim at a reduction of this effort \citep{ryckelynck2005priori}, such that it scales with the dimensionality of the subspace, rather than that of the high-dimensional model. Most hyper-reduction methods define a 'reduced mesh', including a subset of the high-dimensional model's integration points and they can roughly be categorized into the following two classes \citep{grimberg2021mesh, bhattacharyya_hyper-reduction_2025}. So-called 'approximate-then-project' methods evaluate the material law on the reduced mesh, reconstruct the full field stresses \citep[typically via Gappy POD, see][]{everson1995karhunen} and subsequently perform the Galerkin projection. The Discrete Empirical Interpolation Method \citep[DEIM,][]{chaturantabut2010nonlinear} and the Gauss-Newton with Approximated Tensors method \citep[GNAT,][]{carlberg2013gnat} are among the most well-known representatives of this hyper-reduction class. In contrast, 'project-then-approximate' methods first perform the Galerkin projection, and subsequently evaluate the involved integrals on the reduced mesh, which makes them structure preserving. For example, if the underlying model exhibits a potential structure, project-then-approximate methods do not ruin convexity (if present) and tangent symmetry, resulting in an increased robustness compared to approximate-then-project methods \citep[see the overviews by][]{van2018integration, brands2019reduced}. Widespread project-then-approximate hyper-reduction methods include the Energy Conserving Sampling and Weighting method (ECSW) by \citet{farhat2015structure} and the Empirical Cubature Method (ECM) by \citet{hernandez2017dimensional}. The statistically compatible hyper-reduction approach by \citet{wulfinghoff2024statistically} also falls into the 'project-then-approximate' category, but it departs from the necessity to evaluate the microscopic strains (used as input for the material law) at integration points of the underlying high-dimensional model. Instead, they are replaced by a set of strain values, which is at any time representative of the overall microscopic strain distribution and compatible in a statistical sense.\\
It is well known that the selection of an optimal subset of the high-fidelity integration points requires the solution of a computationally intractable minimization problem. Motivated by this observation, integration points differing from the underlying high-dimensional model are identified by the continuous empirical cubature method (CECM) of \citet{hernandez2024cecm}. The method successfully further reduces the number of integration points by allowing {\it all} points of the microstructure as integration point candidates for hyper-reduction.\\
In this work, the Empirically Corrected Cluster Cubature method (E3C) is applied and further extended. Motivated by the encouraging results of the statistically compatible approach mentioned above, the E3C integration points are also detached from the high-fidelity integration points. Originally introduced as solution of a minimization problem \citep{wulfinghoff2025empirically}, recent work \citep{wulfinghoff2026computational} demonstrates that the E3C integration points can be interpreted as approximate solution of an orthogonality condition, ensuring equivalence of the fully integrated ROM and its hyper-reduced counterpart. These orthogonality conditions are unraveled in this work to coincide with the well-known Hill-Mandel condition \citep{hill1963elastic, mandel1972plasticite}, if it is evaluated mode-wise. So far, the method has been successfully applied to magnetostatics \citep{wulfinghoff2025e3c, goldbeck2025computational}, nonlinear elasticity \citep{wulfinghoff2025e3c}, hyperelasticity \citep{wulfinghoff2025homogenization} and crystal plasticity \citep{wulfinghoff2026computational}. In this work, it is formulated for the class of geometrically linear dissipative materials with internal variables and applied to porous and particle-reinforced  elastoviscoplastic microstructures.\\
As a key novelty, the E3C method is combined with an online 'Sequential Subspace Mode Adaptation' procedure to further significantly reduce the number of primary unknowns. The Sequential Subspace Mode Adaptation (SSMA) turns out to fall into the class of sequential subspace optimization methods \citep{narkiss2005sequential}. Starting from a set of modes determined offline, which span a linear subspace, a further reduced affine subspace is determined during the online computation. 
This strategy is motivated from the observation that further reduction of the dimensionality can be achieved, if the employed linear subspaces are replaced by nonlinear manifolds, as has been successfully demonstrated in fluid mechanics \citep[e.g.,][]{barnett2022quadratic, barnett2022neural, geelen2024learning} and solid mechanics \citep[e.g.,][]{scheunemann2025manifold,zhang2026unified}. Typical realizations of such nonlinear manifolds range from quadratic or polynomial ansatz spaces \citep[e.g.,][]{barnett2022quadratic, faust2026nonlinear} to autoencoders \citep[e.g.,][]{lee2020model, fresca2022pod}. These approaches are particularly suited for problems exhibiting strong nonlinearities or moving features such as fronts. In that class of ROMs the solution manifold cannot be efficiently represented by a single global linear subspace, as assumed in the classical approaches outlined above. Instead, the solution exhibits locally varying structures that require more flexible approximation strategies.\\
A similar idea underlies the so-called adaptive ROMs \citep[e.g.,][]{amsallem2012nonlinear, amsallem2015fast, carlberg2015adaptive}, which introduce local reduced bases, representing the key features of the solution or parameter space \citep[e.g.,][]{peng2016nonlinear,daniel2022physics,amsallem2016pebl, faust2024nonlinear, scheunemann2025manifold}. During the online phase, the corresponding local basis is either selected from a set of precomputed affine bases \citep[e.g.,][]{amsallem2016pebl, hatefi2026efficient} or constructed online \linebreak\citep[e.g.,][]{peherstorfer2015online, carlberg2015adaptive}, based on the current state to increase the accuracy of the approximation. In this sense, the present approach can be thought of as a variation of such a locally adaptive approximation, but without the need of explicitly partitioning the space during an offline step.\\
The affine subspace dynamically constructed online by the Sequential Subspace Mode Adaptation presented in this work has the advantages that it does not require the choice of a certain nonlinear manifold class (e.g., quadratic), enables a consistent linearization on the macroscale as well as a fast identification of the affine subspace and can be easily integrated into already implemented pROM frameworks. However, it is not expected, that it allows for a further reduction of integration points compared to the linear subspace model.\\
The remainder of this paper is organized as follows. In Section~\ref{sec:firstorderhomogenization} the first-order homogenization for dissipative microstructures is recalled. Section~\ref{sec:MOR} describes the model order reduction. The E3C hyper-reduction method is introduced in Section~\ref{sec:E3C}, followed by the Sequential Subspace Mode Adaptation in Section~\ref{sec:SSMA_1}. Next, numerical results are shown in Section~\ref{sec:Results} and a conclusion is given in Section~\ref{sec:Conclusion}. \\ 
\textbf{Notation.} Scalar quantities are displayed in light face font, such as \f{a} or \f{A}. Vectors and second-order tensors are represented in boldface font, e.g., \f{\boldsymbol{b}} or \f{\boldsymbol{B}}, while fourth-order tensors are indicated with blackboard boldface letters, e.g., \f{\boldsymbol{\mathbb{C}}}. Macroscopic quantities are denoted with \f{(\bar \bullet)}, while microscopic quantities are written without additional symbol like, e.g., \f{\boldsymbol{u}}. The notations \f{(\tilde \bullet)}, \f{\delta(\bullet)} and \f{(\bullet)_0} mark fluctuations, variations, or a given quantity. At last \f{(\dot\bullet)} denotes the derivative with respect to time and as per usual \f{\T{(\bullet)}} describes the transpose.

\section{First-order homogenization of dissipative microstructures}\label{sec:firstorderhomogenization}
\subsection{Continuum-mechanical framework}
A microscopically heterogeneous dissipative solid is considered and the scale-separation assumption is adopted in order to apply the well-established first-order homogenization framework  \citep[for an overview see][]{schroder2014numerical}. It is assumed that the mechanical small deformation framework summarized in Box~1 defines the micro- and macroscopic problems in addition to the scale transition. The microstructure is assumed to be periodic and composed of \f{\Nph} distinct phases,  characterized by thermodynamically consistent microscopic material models with internal variables, denoted by~\f{\fX} \citep{coleman1967thermodynamics}. This implies phase-dependent constitutive stress relations \f{\fsigma^r(\feps,\fX)} (\f{r=1,\dots,\Nph}) and related evolution equations for the internal variables \f{\dot \fX=\check \fX^r(\feps,\dot \feps,\fX)}. The work at hand focuses on  elastoplasticity as an important subclass of this general framework (for details see Sect.~\ref{sectConst}). However, the model order reduction strategy presented is in principle also applicable to other classes of material models.

\begin{figure}
\centering
\framebox{
\begin{minipage}{.9\textwidth}
\vspace{3mm}
 \begin{itemize}
  \item Macro-problem (neglecting inertia and body forces)
  \begin{itemize}
   \item Strain (kinematics)
   \begin{equation}
    \bar \feps=\fsym{\bgrad{\bar \fu}} \ \rm{in}\  \bar \cB 
   \end{equation}
   \item Equilibrium condition (strong and weak form)
   \begin{equation}
    \bdiv{\bar \fsigma} = \fzero \ \rm{in}\  \bar \cB, \ \ \ \int_{\bar \cB} \bar \fsigma : \delta \bar \feps \d \bar V= \int_{\partial\bar\cB_t}\bar\ft_{0}\cdot\delta\bar\fu\d\bar{A} 
   \end{equation}
   \item Boundary conditions
   \begin{equation}
    \bar \fsigma \bar \fn = \bar\ft_0 \ \rm{on}\ \partial \bar \cB_t,\ \bar \fu = \bar \fu_0\ \rm{on}\ \partial \bar \cB_{\rm{u}}
   \end{equation}
  \end{itemize}
%
  \item Micro-problem
  \begin{itemize}
   \item Kinematics
   \begin{align}
    &\fu(\bx, \fx,t)=\bar \feps(\bx,t) \fx + \tilde \fu(\bx,\fx,t), \  \rm{in}\ \Omega, \ \tilde \fu \ {\rm periodic\ on}\ \partial \Omega \\
    &\feps=\fsym{\grad{\fu}}=\bar \feps+\tilde \feps \ \rm{in}\ \Omega \label{knmtcs}
   \end{align}
   \item Equilibrium condition and antiperiodic tractions (strong and weak form)
   \begin{equation}
    \div{\fsigma} = \fzero \ \rm{in}\  \Omega, \ \ \ \jmp{\fsigma}\fn=\fzero \ \rm{on}\ \partial \Omega, \ \ \ \int_\Omega \fsigma : \delta \tilde \feps \d V=0 \label{microlmb}
   \end{equation}
    + constitutive law (see text)
  \end{itemize}
%
  \item Scale transition
  \begin{flalign}
   &\bar \feps = \langle \feps \rangle, \ \bar \fsigma = \langle \fsigma \rangle \ {\rm with} \ \avg{\bullet}=\frac{1}{\Omega}\int_\Omega (\bullet) \d \Omega&& \label{barqtts}
  \end{flalign}

 \end{itemize}
 \vspace{1mm}
\end{minipage}
}
\caption*{Box~1: Summary of two-scale problem: \f{\fx,\bar\fx}: micro-/ macroscopic position vectors, \f{t}: time, \f{\fu}: displacement, \f{\feps}: strain tensor, \f{\fsigma}: stress tensor, \f{\ft}: traction vector, \f{\fn}: unit normal vector, \f{(\bar \bullet)}: macroscopic quantity, \f{(\tilde \bullet)}: fluctuation, \f{\delta(\bullet)}: virtual quantity, \f{(\bullet)_0}: given quantity, \f{\bar \cB}: solid (macroscale), \f{\Omega}: microstructure. Singular surfaces at phase boundaries are not explicitly discussed, for brevity (but the weak forms hold also in their presence).}
\end{figure}

\subsection{Compatibility}
The microscopic kinematics described by Eq.~\eqref{knmtcs} imply compatibility of the strain field \f{\feps(\fx,t)=\bar\feps+\tilde \feps(\fx,t)}, i.e., there exists a periodic field~\f{\tilde \fu(\fx,t)}, such that~\f{\tilde\feps=\fsym{\grad{\tilde \fu}}}. One can show that a given symmetric second order field~\f{\feps} is compatible if
\begin{equation}
    \int\limits_\Omega \feps : \delta \tilde\fsigma \d \Omega = 0  \label{weakcmpt}
\end{equation}
for all self-equilibrated trial stress fields \f{\delta \tilde\fsigma} (i.e., satisfying \f{\div{\delta \tilde\fsigma}=\fzero} in~\f{\Omega} and~\f{[\![\delta \tilde \fsigma]\!]\fn=\fzero} on~\f{\partial \Omega}) with \f{\avg{\delta \tilde \fsigma}=\fzero}. The weak compatibility condition~\eqref{weakcmpt} can be considered the dual counterpart of the weak equilibrium statement~\eqref{microlmb}\f{_3}. An explanation of Eq.~\eqref{weakcmpt} can be found in \ref{appweakcomp}.

\subsection{Hill-Mandel condition}\label{sec:HM}
In order to ensure energetic consistency between macroscopic and microscopic scales, the Hill-Mandel condition \citep{hill1963elastic, mandel1972plasticite} is imposed as follows:
    \begin{equation}
        \bar \fsigma:\dot{\bar\feps} = \langle \fsigma:\dot\feps\rangle. \label{HM2}
    \end{equation}
If one assumes \f{\bar\fsigma=\langle\fsigma\rangle} and \f{\dot{\bar\feps}=\langle\dot\feps\rangle} to hold, this leads to:
\begin{equation}\label{eq:ortho_fluc}
    \langle\tilde\fsigma:\dot{\tilde\feps}\rangle=0,
\end{equation}
with respect to the microscopic fluctuations of strain \f{\tilde\feps} and stress \f{\tilde\fsigma}. More generally, Eq.~\eqref{eq:ortho_fluc} may be interpreted as orthogonality statement in the sense that the functional spaces of statically admissible stress fluctuations (equilibrium fields) and the kinematically admissible (compatible) strain fluctuations are orthogonal.\\ An alternative representation of the Hill--Mandel condition makes a connection to the principle of virtual work:
\begin{equation}\label{eq:virualwork_HillMandel}
    \bar\fsigma:\delta\bar\feps=\langle\fsigma:\delta\feps\rangle,
\end{equation}
with \f{\delta\bar\feps} and \f{\delta\feps} denoting the variation of macro-/microscopic strain. Using the assumption that \f{\tilde\fu} is periodic, the variation of strain becomes
\begin{equation}\label{eq:kinematic_assumption_HM}
    \delta\feps=\fsym{\grad{\delta\fu}}=\delta\bar\feps+\fsym{\grad{\delta\tilde\fu}}=\delta\bar\feps+\delta\tilde\feps
\end{equation}
with \f{\delta\bar\feps} and \f{\delta\tilde\fu} being arbitrary. Choosing \f{\delta\tilde\fu=\fzero} leads to
\begin{equation}
    \bar\fsigma:\delta\bar\feps=\langle\fsigma:\delta\bar\feps\rangle=\langle\fsigma\rangle:\delta\bar\feps,
\end{equation}
implying
\begin{equation}\label{eq:SigmaFlucVanish}
    \bar\fsigma=\langle\fsigma\rangle{\quad\rm and\quad}\langle\tilde\fsigma\rangle=\langle\fsigma-\langle\fsigma\rangle\rangle=\langle\fsigma-\bar\fsigma\rangle=\fzero.
\end{equation}
Thus, as a consequence of Eq.~\eqref{eq:virualwork_HillMandel} and the kinematic assumption~\eqref{eq:kinematic_assumption_HM}, the macroscopic stress is equal to the volumetric average of the microscopic stress and the microscopic stress fluctuations average out. In addition, choosing \f{\delta\bar\feps=\fzero} results in
\begin{equation}\label{eq:ortho_fluc_variational}
    \langle\fsigma:\delta\tilde\feps\rangle=\langle\tilde\fsigma:\delta\tilde\feps\rangle=0,
\end{equation}
which is nothing but the weak form of the equilibrium condition introduced in Eq.~\eqref{microlmb}.

\subsection{Time discretization}
It is assumed that a suitable time integration algorithm is applied to the evolution equations of the internal variables, such that their value at the end of a given time step \f{\Delta t=t_{n+1}-t_n}, denoted by \f{\fX_{n+1}}, can be formally given for each phase~\f{r} by
\begin{equation}
    \fX_{n+1}=\fX^r(\feps_{n+1},\fX_n), \label{Xalgo}
\end{equation}
where the index '\f{n+1}' will be neglected in the following. If not mentioned otherwise, the subsequent sections concentrate on the time discrete model.

\section{Model order reduction}\label{sec:MOR}
\subsection{Galerkin projection}\label{sec:MOR_1}
\subsubsection{Discretization in terms of strain modes}
Projection-based reduced order models approximate the microscopic strain fluctuation field \f{\tilde \feps(\fx)=\feps(\fx)-\bar \feps} at time \f{t_{n+1}} by a linear combination of \f{\Nmd} strain fluctuation modes \f{\tcEps_k(\fx)}, representing global shape functions, such that
\begin{equation}
    \feps(\fx)=\bar \feps+\tilde \feps(\fx)\approx\bar\feps+\sum_{k=1}^{N_{\rm md}}\xi_{k}\tcfE_{k}(\fx) \label{epsdiscr}
\end{equation}
holds, with \f{\xi_{k}} denoting the mode coefficients. The virtual strain fluctuations \f{\delta \tilde\feps(\fx)} are discretized in analogy.
Here, the dependence on the macroscopic position~\f{\bx} has been neglected for the sake of notational simplicity.
The origin of the modes\footnote{In this contribution, the modes are obtained through singular value decomposition.} is of negligible relevance for the concepts developed in the following discussion (see the Introduction for further reference).
\subsubsection{Galerkin projection and solution of the microscopic problem}
The reduced form of the microscopic equilibrium conditions is obtained by insertion of approximation~\eqref{epsdiscr} into the weak form~\eqref{microlmb}:
\begin{equation}
    \frac{1}{\Omega}\int_{\Omega}\fsigma:\delta\tilde\feps{\rm d}V=\frac{1}{\Omega}\int_{\Omega}\fsigma:\biggl(\sum_{k=1}^{N_{\rm md}}\delta\xi_{k}\tcfE_{k}(\fx)\biggr){\rm d}V=\sum_{k=1}^{N_{\rm md}}\underbrace{\biggl(\frac{1}{\Omega}\int_{\Omega}\fsigma:\tcfE_{k}(\fx){\rm d}V\biggr)}_{R_{k}=0}\delta\xi_{k}. \label{rsdls}
\end{equation}
In this context, the residuals \f{\fR(\fxi)=\T{(R_1, \dots, R_{\Nmd})}} form a set of nonlinear equations for the unknown mode coefficients, that can be solved via Newton's method. The solution \linebreak\f{\fxi=\T{(\xi_1,\dots,\xi_{\Nmd})}} delivers the microscopic strain distribution via Eq.~\eqref{epsdiscr}, the internal variables via Eq.~\eqref{Xalgo} and the stresses via the constitutive functions \f{\fsigma^r(\feps,\fX)}. Finally, the macroscopic stress~\f{\bar \fsigma} is obtained by averaging (Eq.~\eqref{barqtts}).\\
{\bf Remark.} In analogy to the strains, the stress fluctuations may also be assumed to occupy a low-dimensional subspace, spanned by stress fluctuation modes \f{\tilde \fS_m(\fx)}:
\begin{equation}
     \fsigma(\fx)=\bar \fsigma + \tilde \fsigma(\fx)
     =\bar \fsigma + \sum\limits_{m=1}^{\Nmdsgma} \mu_m \tilde \fS_m(\fx). \label{strssmds1}
\end{equation}
Galerkin-projecting the weak compatibility condition~\eqref{weakcmpt} onto this reduced subspace yields 
\begin{equation}
    \frac{1}{\Omega} \int \limits_\Omega \feps : \tilde \fS_m \d \Omega = 0. \label{cmptGlrkn}
\end{equation}

\subsubsection{Specification for generalized standard materials}\label{sec:GSM}
For generalized standard materials \citep{halphen1975materiaux}, the solution of the micro-problem, outlined in Box~1, minimizes a time-incremental pseudo-elastic potential, which under certain assumptions takes the form \citep{ortiz99, miehe02}:
    \begin{equation}
        \bar\pi_{\Delta}(\bar\feps)=\infi{\tilde {\blds u} \#} \langle \pi_\Delta (\fx, \feps) \rangle \ \ \ {\rm with} \
        \pi_\Delta (\fx, \feps) = \infi{\blds X} \left( \psi(\fx,\feps, \fX) + \Delta t \phi(\fx, \fX) \right), \label{infiu}
    \end{equation}
    where '\#' marks the periodicity of \f{\tilde \fu}, \f{\psi} is the free energy density, and~\f{\phi} is the dissipation potential. Alternatively, the infimum may be searched for within the space of compatible (i.e., kinematically admissible) strain fluctuations. If it is assumed that the strain fluctuation field resides in the linear subspace spanned by the modes \f{\tcEps_k}, it is sufficient to restrict the minimization to that lower-dimensional space:
    \begin{equation}
        \infi{\tilde {\blds u} \#} \langle \pi_\Delta (\fx, \feps) \rangle
        =\infi{\blds \xi} \big\langle \pi_\Delta \big(\fx, \feps(\bar \feps, \fxi) \big) \big\rangle \ \ \ 
        {\rm with} \ \feps(\bar \feps, \fxi) = \bar \feps + \sum\limits_{k=1}^{\Nmd} \xi_k \tcEps_k. \label{infixi}
    \end{equation}
    Under this assumption, Eq.~\eqref{rsdls} is interpreted as stationarity condition of the minimization problem \eqref{infixi} and delivers the exact solution. Otherwise it is an approximation.
    
\subsubsection{Mode-restricted Hill-Mandel condition}\label{sec:HM_MOR}
The variational form of the Hill-Mandel condition given in Eq.~\eqref{eq:virualwork_HillMandel} can be naturally transferred  the reduced order settings. The fluctuation fields will therefore be approximated in suitable reduced spaces, as shown in the previous section. In particular, the stress fluctuation modes \f{\tilde \fS_m} can be obtained as a linear combination of a sufficiently rich set of microscopic stress fluctuation snapshots as
\begin{equation}\label{eq:stress_modes}
    \tilde \fS_m(\fx) = \sum\limits_{n} a_{mn} \tilde \fsigma_n(\fx).
\end{equation}
Due to the static admissibility of the snapshots \f{\tilde\fsigma_n}, the corresponding modes \f{\tilde \fS_m} remain statically admissible (i.e., they satisfy Eq.~\eqref{microlmb}). Combining Eq.~\eqref{epsdiscr} with the equilibrium condition~\eqref{eq:ortho_fluc_variational} yields
\begin{equation}
     \langle \tilde \fsigma:  \delta\tilde\feps\rangle = \sum_l \,
    \langle \tilde \fsigma : \tcEps_l \rangle \delta \xi_l. \label{HM1}
\end{equation}
The arbitrariness of \f{\delta\xi_l} implies \f{\langle\tilde\fsigma:\tcEps_l\rangle=0}. Further, multiplication with \f{a_{kn}} and summing over \f{n} (see Eq.~\eqref{eq:stress_modes}) delivers
\begin{equation}
    \langle \tilde \fS_k : \tcEps_l \rangle = 0. \label{HMdiscr}
\end{equation}
This relation can be interpreted as an orthogonality condition between the reduced spaces spanned by the stress and strain fluctuation modes. It is the mode-restricted version of the orthogonality statement in Eq.~\eqref{eq:ortho_fluc}. Equation \eqref{HMdiscr} can thus be understood as reduced-order representation of the Hill-Mandel condition, restricted to the linear subspaces spanned by the fluctuation modes.

\section{E3C Hyper-reduction}\label{sec:E3C}
\subsection{General numerical integration procedure}
The numerical computation of the residuals \f{R_k} (Eq.~\eqref{rsdls}) and macroscopic stress (Eq.~\eqref{barqtts}) typically requires an integration/cubature scheme (not necessarily hyper-reduced) of the form
\begin{equation}
    R_k=\frac{1}{\Omega}\sum\limits_{q=1}^{\NipFE} \fsigma^q(\feps^q,\fX_n):\tcEps_k^q \, \Omega^q=0, \ \ \ 
    \avg{\fsigma} = \frac{1}{\Omega}\sum\limits_{q=1}^{\NipFE} \fsigma^q(\feps^q,\fX_n) \Omega^q  \label{rsdlqdrtre}
\end{equation}
with the number of integration points~\f{\NipFE}, integration domains/weights \f{\Omega^q} and 
\begin{equation}
    \feps^q=\bar \feps+\sum\limits_{l=1}^{\Nmd} \xi_l \tcEps_l^q.
\end{equation}
The averaging operator correspondingly now reads
\begin{equation}\label{eq:averaging_operator}
    \avg{\bullet} = \frac{1}{\Omega} \sum\limits_{q=1}^{\NipFE} (\bullet)^q \Omega^q 
\end{equation}
throughout Section~\ref{sec:E3C}.
The \f{\tcEps_k^q} (\f{k=1,\dots,\Nmd}) and~\f{\Omega^q} are usually obtained from a high-fidelity model by evaluation at the integration points~\f{\fx^q}, i.e., in that case \f{\tcEps_k^q=\tcEps_k(\fx^q)}.\\
The continuous modes \f{\tcEps(\fx)} are thus replaced by the integration point-wise resolved modes~\f{\tcEps_k}, which are collected in the matrix of strain fluctuation modes, \f{\tcEps}, as follows (assuming Mandel notation\footnote{Mandel notation: \f{\fsigma=\T{(\sigma_{11},\sigma_{22},\sigma_{33},\sqrt{2}\sigma_{23},\sqrt{2}\sigma_{13},\sqrt{2}\sigma_{12})}}.}):
\begin{equation}\label{eq:matrixofstrainfluctmodes}
    \tcEps_k=\begin{pmatrix}
        \tcEps_k^1 \\ \vdots \\ \tcEps_k^{\NipFE}
    \end{pmatrix} \  \ \ 
    \Rightarrow \ \tcEps=\begin{pmatrix}
        \tcEps_1 & \dots & \tcEps_{\Nmd}
    \end{pmatrix} \ \ \ 
    \Rightarrow \tilde \fE = \begin{pmatrix}
        \tilde\feps^1\\ \vdots \\ \tilde \feps^{\NipFE}
    \end{pmatrix}=\tcEps\fxi.
\end{equation}
Thus, \f{\tcEps_k} is the discrete counterpart of \f{\tcEps_k(\fx)}. The two quantities are uniquely distinguished by consistently keeping the position-dependence in the notation of the latter. The strain fluctuation modes are constrained to vanish on average:
\begin{equation}
    \avg{\tcEps_k}=\frac{1}{\Omega}\sum\limits_{q=1}^{\NipFE}\tcEps_k^q \Omega^q = \fzero, \label{mdeavg}
\end{equation}
ensuring that~\f{\avg{\tilde \feps}=\fzero}.\\
For later use, the matrix of macroscopic strain modes, \f{\bcEps}, is also defined:
\begin{equation}
    \bcEps_m=\begin{pmatrix}
        \bcEps_m^1 \\ \vdots \\ \bcEps_m^{\NipFE}
    \end{pmatrix} \  \ \ 
    \Rightarrow \ \bcEps=\begin{pmatrix}
        \bcEps_1 & \dots & \bcEps_6
    \end{pmatrix} \ \ \ 
    \Rightarrow \fE = \begin{pmatrix}
        \feps^1\\ \vdots \\  \feps^{\NipFE}
    \end{pmatrix}=\bcEps \bar \feps + \tcEps\fxi \label{Eeq}
\end{equation}
with~\f{\bcEps_1^q=\T{(1,0,0,0,0,0)}}, \dots, \f{\bcEps^q_6=\T{(0,0,0,0,0,1)}} (\f{q=1,\dots,\NipFE}).\\
{\bf Remark.} The statement that the~\f{\tcEps_k^q} and integration domains~\f{\Omega^q} {\it must} originate from the discretization of a high fidelity model, can be relaxed by considering the following: One and the same microstructure can be discretized in an infinite number of ways, i.e. the~\f{\tcEps_k^q} are not unique. However, all discretizations lead (if they are consistent with the continuum model) to essentially the same results, if a sufficient numerical resolution can be assumed. This approach will be used in the following for the definition of the E3C modes.

\subsection{E3C Hyper-reduction}\label{sectorth}
In the sequel, a hyper-reduced model is considered, i.e., the number~\f{\NipFE} of integration points is significantly smaller than that of the high-fidelity model \f{\NipFE\ll\NipFE^{\rm ref}}.\\
In order to derive hyper-reduced modes \f{\tcEps_k^{\rm T}=(\tcEps_k^{1{\rm T}},\dots,\tcEps_k^{{\NipFE}{\rm T}})^{\rm T}} (see Eq.~\eqref{eq:matrixofstrainfluctmodes}), the hyper-reduced model is required to satisfy the Hill-Mandel condition (as introduced in Eq.~\eqref{eq:virualwork_HillMandel})
\begin{equation}
    \bar \fsigma : \delta\bar \feps = \avg{\fsigma:\delta\feps} = \frac{1}{\Omega} \sum\limits_{q=1}^{\NipFE} \fsigma^q:\delta\feps^q \Omega^q \label{HMHR}
\end{equation}
for {\it any} deformation process \f{\{\bar \fsigma(t), \bar\feps(t), \fxi(t)\}} (or its time-discrete counterpart). Here, \f{\bar \fsigma} (left hand side) and \f{\fxi} are understood as the high-fidelity solution\footnote{If the modes~\f{\tcEps^{\rm FE}_k} can be assumed orthonormal (i.e., obtained from a truncated singular value decomposition of finite element snapshot data), then any mode coefficient~\f{\xi_k} (\f{k\in\{1,\dots,\Nmd\}}) can be directly deduced from a given finite element snapshot \f{\fE^{\rm FE}} (compare Eq.~\eqref{Eeq}) via projection, i.e., \f{\xi_k=\tcEps^{\rm FE}_k \cdot \fE^{\rm FE}} \citep{barnett2022quadratic}.}, while the stress~\f{\fsigma^q} at generalized integration point~\f{q} of the hyper-reduced model (right hand side) is obtained by evaluation of the material law \f{\fsigma^{q}(\feps^{q},\fX_{n}^{q})} with input strain
\begin{equation}
    \feps^q=\bar\feps+\sum\limits_{k=1}^{\Nmd} \xi_k\tcEps_k^q \label{eq:strainatIPs}
\end{equation}
with both \f{\bar \feps} {\it and}~\f{\xi_k} being given by the high-fidelity solution. By construction \f{\avg{\feps}=\bar\feps} holds (due to the constraint applied in Eq.~\eqref{mdeavg}), such that
\begin{equation}
    \bar \fsigma :\delta\bar \feps = \bar\fsigma : \avg{\delta\feps} = \avg{\fsigma:\delta\feps} \ \ \
    \Leftrightarrow \ \ \ \avg{(\fsigma-\bar\fsigma):\delta\feps} = 0
\end{equation}
with averaging operator as introduced in  Eq.~\eqref{eq:averaging_operator} results. Including the decomposition of the strain into macroscopic and fluctuation parts, i.e., \f{\feps^q=\bar\feps+\tilde\feps^q} (compare Eq.~\eqref{knmtcs}), and considering the arbitrariness of~\f{\delta\bar\feps}, the following two equations are obtained
\begin{equation}
     \avg{(\fsigma-\bar\fsigma):\delta\tilde\feps} = 0, \ \ \ \avg{\fsigma-\bar\fsigma}=\fzero. \label{HMsep}
\end{equation}
A comparison with Eqns.~\eqref{eq:SigmaFlucVanish} and \eqref{eq:ortho_fluc_variational} shows that the hyper-reduced model is consistent with the continuous case.
It is noted that the corresponding stress fluctuations \f{\tilde \fsigma^q:=\fsigma^q-\bar\fsigma} are defined with respect to the macroscopic stresses of the high-fidelity model. As a crucial observation, one can expect these stress fluctuations to reside in a low dimensional subspace, since the strain fluctuations \f{\tilde \feps^q=\tcEps\fxi} exhibit the very same property:
\begin{equation}
    \begin{pmatrix}
        \tilde\fsigma^1\\ \vdots \\ \tilde\fsigma^{\NipFE} 
    \end{pmatrix}=\sum\limits_{m=1}^{\Nmdsgma} \mu_m \tilde \fS_m, \label{strssmds}
\end{equation}
where the \f{\tilde \fS_m \in \ffR^{6\NipFE}} are not to be confused with the continuous modes \f{\tilde \fS_m(\fx)} in Eq.~\eqref{strssmds1}. It is recalled that the strains~\f{\fE=(\feps^{1{\rm T}}, \dots, \feps^{\NipFE{\rm T}})^{\rm T}=\bcEps\bar \feps+\tcEps\fxi} are given in terms of prescribed macro-strains \f{\bar\feps} {\it and} prescribed mode coefficients \f{\fxi}. Furthermore, it is noted that stresses~\f{ \fsigma^q(\feps^q,\fX_n)} -- and consequentially also the modes~\f{\tilde \fS_m} -- are functions of the strain fluctuation modes~\f{\tcEps_k} (trough Eq. \eqref{eq:strainatIPs}).\\
The Hill-Mandel condition in the form~\eqref{HMsep} holds for arbitrary processes \f{\{\bar\fsigma(t), \bar\feps(t), \fxi(t)\}}, if
\begin{equation}
    \avg{\tilde\fS_m:\tcEps_n}=0, \ \  \ \avg{\tilde\fS_m}=\fzero, \label{orthconds}
\end{equation}
since in this case
\begin{equation}
    \left\langle \sum_m\mu_m\tilde\fS_m:\sum_n\delta\xi_n\tcEps_n \right\rangle=\avg{\tilde\fsigma:\delta\tilde\feps}=0, \ \  \ \left\langle \sum_m \mu_m \tilde\fS_m\right\rangle=\avg{\tilde\fsigma}=\fzero. \label{eq:HM_Contributions}
\end{equation}    
This is in analogy to the continuous version introduced in Sec.~\ref{sec:HM} in Eq.~\eqref{eq:SigmaFlucVanish} and the mode specific versions in Sec.~\ref{sec:HM_MOR} in Eq.~\eqref{HMdiscr}.
Hyper-reduced strain fluctuation modes, which result in satisfaction of the hyper-reduced and mode-restricted Hill-Mandel conditions \eqref{orthconds}, have the following implications:
\begin{enumerate}
    \item They guarantee equivalence with the high-fidelity model, since for a given history of mode coefficients \f{\fxi} (taken form the high-fidelity model) multiplication of Eq.~\eqref{orthconds} with \f{\mu_m}, summation over~\f{m} and division by~\f{\Omega} yields \f{R_k=0} and \f{\bar \fsigma=\avg{ \fsigma}} (Eq.~\eqref{rsdlqdrtre}). Here, it has been used that there are always coefficients \f{\mu_m}, such that the stress fluctuations can be exactly represented as linear combination of the modes \f{\tilde\fS^{q}_{m}}.
    \item The reduced-order model satisfies hyper-reduced (and Galerkin-projected) versions of 
    \begin{itemize}
        \item the equilibrium condition~\eqref{rsdlqdrtre}\f{_1}:
        \begin{equation}
          \langle\fsigma:\tcEps_{k}\rangle=0,
         \end{equation}    
        as is already clear from point 1.
        \item the compatibility condition (compare Eq.~\eqref{cmptGlrkn}):
    \begin{equation}
        \langle\feps:\tilde\fS_m\rangle=0,
    \end{equation}
    which follows from multiplication of Eq.~\eqref{orthconds}\f{_1} by~\f{\xi_k}, summation over~\f{k} and division by~\f{\Omega}. In addition, use has been made of the property~\f{\bar\feps:\avg{\tilde \fS_m}=\fzero}.    
   \end{itemize}
\end{enumerate}
Points 1 and 2 hold under the assumption that both stress and strain fluctuations reside in a finite-dimensional subspace.

\subsection{Computation of the E3C modes}
The solution of Eqns.~\eqref{orthconds}\f{_1} and \eqref{orthconds}\f{_2} is complicated by the fact that the \f{\tilde\fS_m} are functions of the~\f{\tcEps_n}, and these functions are not explicitly given. For this reason, the stress fluctuation modes are replaced by a sufficiently large number of exemplary stress fluctuations. For this purpose, Eqns.~\eqref{orthconds} are rearranged once again by multiplication with~\f{\mu_m}, summation over~\f{m} and division by~\f{\Omega}, leading to~\f{R_k=0} and~\f{\avg{\fsigma}-\bar\fsigma=\fzero}. Squaring, summation over~\f{k}, multiplication by~\f{1/2} and repeating this procedure for \f{\Nsim} simulations, yields
\begin{equation}
    c(\tcEps)=\frac{1}{2}\sum\limits_{s=1}^{\Nsim}\sum\limits_{n=0}^{N_t^s-1} \left[
    a \sum_{k=1}^{\Nmd} \underbrace{ 
    \langle\fsigma^{s}_{n+1}:\tcEps_k\rangle
    ^2 }_{(R^s_{k(n+1)})^2}
    +\left\|\langle\fsigma^s_{n+1}\rangle-\bar\fsigma^{s}_{n+1}\right\|^2\right]=0.\label{cstfctn}
\end{equation}
Here, \f{N_t^s} is the number of time steps of simulation~\f{s\in\{1,\dots,\Nsim\}}. In addition, the strains \f{\feps^{qs}_{n+1}=\bar\feps^s_{n+1}+\tcEps\fxi^s_{n+1}} are calculated using the {\it given} macroscopic strains and the mode coefficients taken from the high-fidelity model. The microscopic stresses are given by the functions/algorithms \f{\fsigma^{qs}_{n+1}=\fsigma^q(\feps^{qs}_{n+1},\fX^{qs}_n)}.\\
Equation \eqref{cstfctn} is expected to be equivalent to the hyper-reduced and mode-restricted conditions~\eqref{orthconds} (and therefore to the Hill-Mandel condition~\eqref{HMHR}), if the set of prescribed macroscopic strains~\f{\bar \feps^s_{n+1}} is sufficiently rich.
The expression~\f{c(\tcEps)} in Eq.~\eqref{cstfctn} allows for the approximate determination of the desired E3C modes\footnote{The initial set of modes is generated through k-means clustering. Compare with \cite{goldbeck2025computational} for general insights regarding the approach.}~\f{\tcEps}, if it is interpreted as a cost function. The minimization of the latter is thus nothing but the minimization of the error in Eq.~\eqref{cstfctn} (right hand side zero). The first contribution ensures that the model learns the correct micro-equilibrium states (\f{R_k\approx 0}) and the second contribution enforces a meaningful macro-stress prediction (\f{\avg{\fsigma}\approx\bar\fsigma}).  The user-defined weighting factor~\f{a} ensures that the two contributions of the cost function are of the same order of magnitude. The exact satisfaction of Eqns.~\eqref{orthconds}, which would theoretically require a certain minimum number of integration points~\f{\NipFE} to prevent the equation system from being underdetermined, is thus not necessary. Instead, smaller integration point numbers can also be used, if larger errors are tolerable.\\
In this work, the cost function~\eqref{cstfctn} is minimized by using the Levenberg Marquardt algorithm. The required cost function derivatives can be found in \ref{APPCostfunctionMinimizationDerivatives}. For a more extensive discussion refer to \citet{wulfinghoff2026computational}. As a novel feature, an additional volumetric damping term has been included, i.e., the damping of the spherical part of the modes~\f{\tcEps_k^q} differs from that of the deviatoric part, motivated by the observation that the volumetric strains are (on average) significantly smaller, due to the purely elastic nature of the volumetric material response. Details can be found in \ref{appdamping}.

\section{Sequential Subspace Mode Adaptation}\label{sec:SSMA_1}
\subsection{Motivation}
The offline training simulations for snapshot collection may be considered 'complete', if the resulting set of strain fluctuation modes \f{\tcEps_k(\fx)} (Eq.~\eqref{epsdiscr}) spans exactly that reduced subspace of all (kinematically admissible) strain fluctuation fields~\f{\tilde\feps(\fx)}, which is relevant to simulate largely \textit{arbitrary} deformations of the specific microstructure at hand. However, a \textit{specific} solution may not require the full wealth of possible deformations encoded within the modes, and it may be beneficial to construct a 'reduced subspace of the reduced subspace'. The 'Sequential Subspace Mode Adaptation' (SSMA), presented in the following, performs such a second compression step online based on a given set of modes.\\
The SSMA method is independent of the hyper-reduction approach and so is the subsequent presentation. It is meant to be easily integrated into existing implementations and does not require any further offline computations. With this being an advantage over nonlinear manifold-based methods, it should be mentioned that the SSMA, however, does not enable a further reduction of hyper-reduced integration points, and preserves those of the linear subspace-based model.\\
For illustration purposes, the following presentation is restricted to generalized standard materials with incremental potential structure (see Eqns.~\eqref{infiu} and~\eqref{infixi}). The SSMA method \textit{may} also work for more general material models. 

\subsection{Sequential Subspace Mode Adaptation Algorithm}\label{sectadaptmodes}
\begin{figure}[h]
\centering
    \psfrag{mini}[tl][c]{minimum on affine subspace}
    \psfrag{glob}[cl][ct]{global minimum}
  \psfrag{grd}[c][c]{\hspace{-1.5mm}\f{-\pd{\bar \pi_\Delta}{\fxi}}}
  \psfrag{cont}[bc][c]{contour lines of \f{\bar \pi_\Delta}}
  \psfrag{aff}[lc][c]{affine subspace}
  \psfrag{xi0}[c][c]{\f{\fxi_0}}
  \psfrag{xi1}[c][c]{\f{\xi_1}}
  \psfrag{xi2}[c][c]{\f{\xi_2}}
  \psfrag{xi3}[c][c]{\f{\xi_3}}
  \psfrag{H}[c][c]{\f{\|\fH\|} [A/mm]}
  \includegraphics[width=0.5\textwidth]{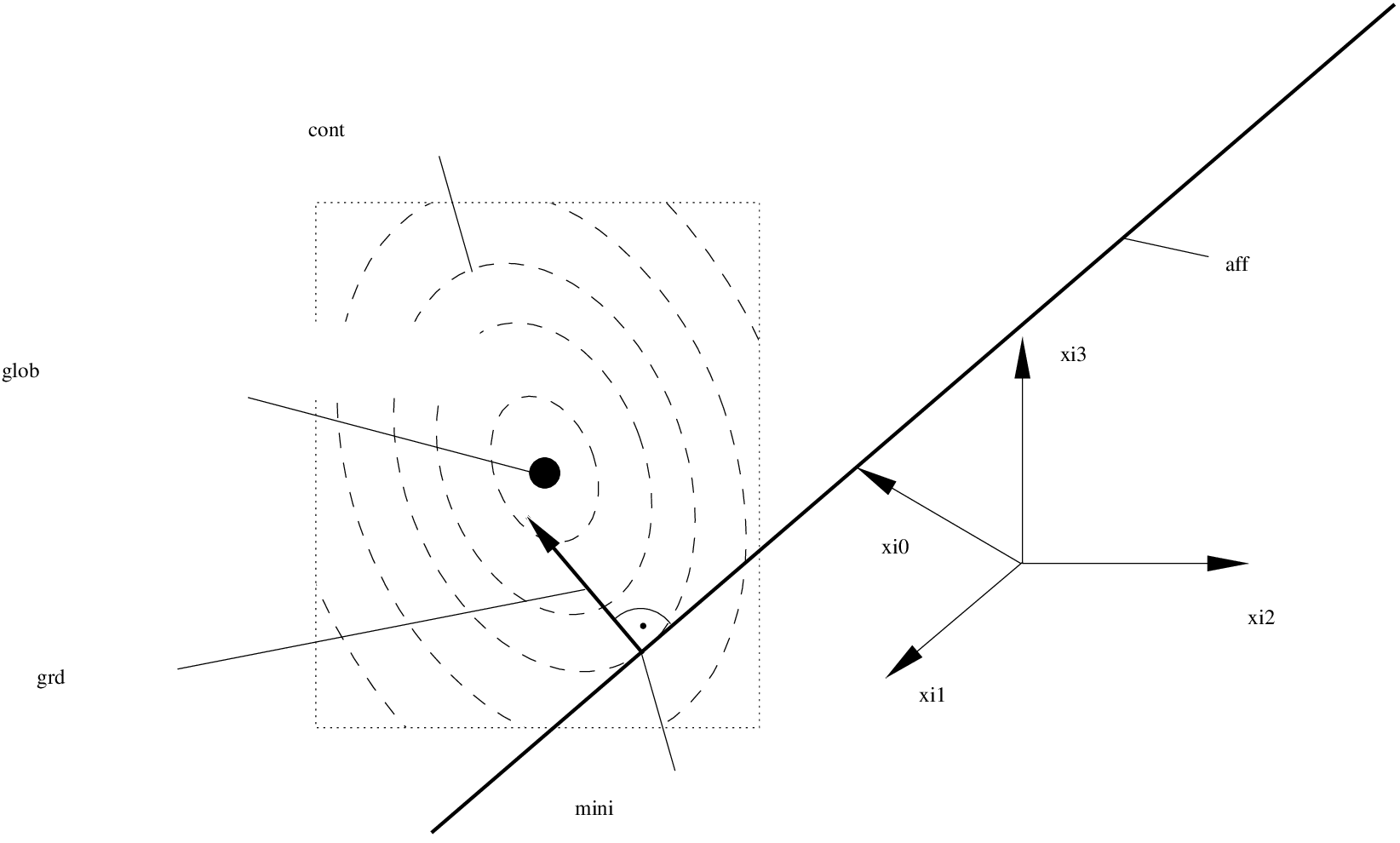}
\caption{Sketch of the affine subspace extension in a simplified three-dimensional setting. The contour lines of \f{\bar \pi_\Delta} (dashed lines) are depicted in a plane parallel to the \f{\xi_2}-\f{\xi_3}-plane (dotted rectangle). The initial affine subspace is here represented as a straight line in \f{\xi_1}-direction. At the minimum of the potential~\f{\bar \pi_\Delta} along that line, the gradient \f{\partial \bar \pi_\Delta/\partial \fxi} serves as new direction for the affine subspace extension, which would in this case correspond to the plane spanned by the gradient and the initial affine subspace. Subsequently, the minimum within this plane is sought for.}
\label{manifoldappr}
\end{figure}

The Sequential Subspace Mode Adaptation procedure seeks an approximate solution within a low-dimensional {\it affine} subspace. The affine subspace is formally defined in terms of the mapping
\begin{equation}
    \fxi(\feta)= \fxi_0 + \fA \feta = \fxi_0 + \pd{\fxi}{\feta} \feta
\end{equation}
with the reduced coordinates \f{\feta \in \ffR^{\Nmd^0}}, constant matrix \f{\fA=\partial \fxi/\partial \feta} and \f{\Nmd^0<\Nmd}. The corresponding strain field follows as:
\begin{equation}
    \feps(\fx,\bar\feps,\feta) = \bar \feps + \sum\limits_{k=1}^{\Nmd} \xi_k(\feta) \tcEps_k(\fx).
\end{equation}
The microscopic equilibrium state in terms of the mode coefficients~\f{\fxi} is thus approximated as a minimum of the potential \f{\bar \pi_\Delta(\bar \feps, \fxi)} (Eq.~\eqref{infixi}) restricted to this affine subspace. If the latter captures the deformation patterns adopted by the high-fidelity model with sufficient accuracy, then the potential minimum in this subspace is expected to be close to the global one (see Fig.~\ref{manifoldappr}). At this solution, the potential gradient \f{\partial \bar \pi_\Delta/\partial \fxi=\fR(\fxi(\feta))} is orthogonal to the affine subspace, which can be seen as follows. Using \f{\partial \bar \pi_\Delta/\partial \fxi=\fR}, the necessary condition for a minimum reads
\begin{equation}
    \d \bar \pi_\Delta =\pd{\bar \pi_\Delta}{\feta}\cdot \d \feta = \pd{\bar \pi_\Delta }{\fxi} \cdot \pd{\fxi}{\feta} \d \feta = \fR \cdot \d \fxi = 0
\end{equation}
for all \f{\d \fxi} within the affine subspace. This observation has an important consequence: the residual \f{\fR} provides a direction that is not yet contained in the current affine subspace. Therefore, it contains information which can be used to enrich the current affine subspace (roughly) in the direction of the global minimum (see Fig.~\ref{manifoldappr}).
\\
Whenever the affine subspace minimum deviates too much from the global minimum, the gradient thus naturally qualifies as a candidate to extend the set of modes:
\begin{equation}
    (\tcEps_1(\fx), \dots, \tcEps_{\Nmd^0}(\fx)) \ \leftarrow \ \left(\tcEps_1(\fx), \dots ,\tcEps_{\Nmd^0}(\fx), \tcEps(\fx) \fR \right) \ \ \ {\rm with} \ \ \ 
    \tcEps(\fx)=(\tcEps_1(\fx), \dots, \tcEps_{\Nmd}(\fx)). \label{mdeextnsn}
\end{equation}
Out of all of the system's (inactive) degrees of freedom, it thus chooses that degree (or rather linear combination thereof), which instantaneously promises the maximum decrease in potential. The residual can thus be interpreted as capturing missing deformation patterns that are required to further decrease the potential. A repeated extension of the affine subspace will eventually lead to a good approximation of the solution.\\
The leading modes are dominant and can be interpreted as deformation patterns shared by a significant fraction of the microscopic deformation states seen during the training stage using a high-fidelity solution. As a consequence, even the reduced model based exclusively on the leading modes often yields a significant energy reduction compared to the homogeneous strain state. 
Motivated by these observations, the following Sequential Subspace Mode Adaptation procedure is proposed:
\begin{enumerate}
    \item For given \f{\bar \feps} and given state \f{\fxi_n} at time~\f{t_n}, seek an approximate solution at~\f{t_{n+1}} in terms of the first \f{\Nmd^0} mode coefficients, i.e., set
    \begin{equation}
        \fxi(\feta) = \fxi_n+\pd{\fxi}{\feta}\feta, \ \ \ 
        \pd{\fxi}{\feta}=\begin{pmatrix}
            \fI \\ \fzero
        \end{pmatrix}=
        \begin{pmatrix}
            1&0&\dots&0\\
            0&1&\dots&0\\
            \vdots&\vdots&\ddots&\vdots\\
            0&0&\dots&1\\
            0&0&\dots&0\\
            \vdots&\vdots&\ddots&\vdots\\
            0&0&\dots&0
        \end{pmatrix} \in \ffR^{\Nmd \times \Nmd^0},
    \end{equation}
    that is
    \begin{equation}
        \xi_i(\feta) =
        \begin{cases}
            \xi_{i n} + \eta_i, & i = 1,\dots,\Nmd^0, \\
            \xi_{i n}, & i = \Nmd^0+1,\dots,\Nmd.
        \end{cases}
    \end{equation}
    
    \item Project the nonlinear system of equations~\eqref{rsdls} onto the affine subspace, i.e., find the solution~\f{\feta} to
    \begin{equation}
      \fR_\eta=   \T{\left(\pd{\fxi}{\feta}\right)} \fR(\fxi(\feta)) = \fzero. \label{redrsdl}
    \end{equation}
    If the dimension of \f{\feta} reached some user-defined value \f{\rightarrow} {\tt stop}.
    \item At the solution \f{\feta} of the reduced equation system~\eqref{redrsdl}, increase the dimension of \f{\feta}, i.e., of the affine subspace by one by adding the residual \f{\fR(\fxi(\feta))} as additional column to \f{\partial \fxi/\partial \feta}.
    Go to 2.
\end{enumerate}
The proposed method can be categorized as sequential subspace optimization (SESOP) algorithm \citep{narkiss2005sequential}. This class of algorithms exhibits pronounced similarity with the conjugate gradient (CG) method. In the present computational homogenization context, the advantage of the proposed approach over CG is that it enables (under certain conditions) a consistent linearization of the micro-model with respect to the macro-strain~\f{\bar \feps}, being a necessary requirement for superior performance of the two-scale model.
\subsection{Implementation aspects}
From a performance point of view, it may be disadvantageous to update the affine subspace at a given macroscopic integration point each time, the microscopic problem is solved. Instead, it is proposed to perform this update only once per time step according to the following strategy. First, a fixed number (typically 3 or 4) macroscopic global Newton iterations are carried out, solving the microscopic problem on the affine subspace determined during the previous time step. Then, the affine subspace is updated according to the approach described above and the remaining Newton iterations are performed with fixed affine subspace, thus enabling a consistent linearization and quadratic convergence. This implies that, each time the update is performed, the active modes are reset to the first \f{\Nmd^0} modes (see Sect.~\ref{sec:Results_ConvergencEval} and \ref{app:twoscalealgo} for additional details). Furthermore, it is mentioned that the order of the matrix multiplications for the local assembly of the tangent~\f{\partial \fR_\eta/\partial \feta} (see Eq.~\eqref{redrsdl}) may have a decisive effect on the overall performance.

\section{Results}\label{sec:Results}
\subsection{Snapshot collection through Latin Hypercube Sampling}\label{Sect:Snapshots}
In order to apply the methods introduced in the previous sections, a set of snapshots is generated from high-fidelity simulations, here based on the finite element method (FEM). The corresponding load path directions are constructed using a Latin Hypercube Sampling (LHS) strategy \citep[compare, e.g.,][]{McKay1979_LHS} in strain space to ensure a broad and well-distributed coverage of possible deformation states. Monotonic loading is assumed. Depending on the microstructure, the sampling procedure is slightly adapted. For porous materials, the full strain space is considered, allowing for volumetric deformations. In contrast, for non-porous microstructures, sampling is kept close to the deviatoric subspace by choosing a comparably smaller random value for the trace of the strain tensor, such that the pressure is in the order of the stress deviator magnitude. This is achieved by constructing an explicit basis of the deviatoric strain space and performing the sampling in the corresponding reduced coordinate system. The Latin Hypercube Sampling is applied in this reduced space, resulting in a set of well-distributed sample points within this hypercube (edge length \f{2}). To obtain a more isotropic distribution of strain states, the samples are subsequently restricted to the unit hypersphere (hyperradius \f{1}) by rejecting points with a norm larger than unity. Points from within the hypersphere are projected to its surface.

\subsection{Constitutive equations and hardening laws}\label{sectConst}
All subsequently discussed simulations employ a standard elastoviscoplastic material law with power law overstress type flow rule and isotropic hardening:
\begin{equation}
    \feps=\fepse+\fepsp, \ \ \ \fsigma=\ffC:\fepse, \ \ \ 
    \dot \feps^{\rm p}=\sqrt{\frac{2}{3}} \dot \varepsilon_0 \left\langle \frac{\sigmavM-\sigmay}{\sigma^{\rm D}} \right\rangle^p \frac{\fsigma'}{\|\fsigma'\|} \label{plstmdl}
\end{equation}
with elastic strain \f{\fepse}, plastic strain \f{\fepsp}, isotropic stiffness tensor~\f{\ffC}, stress deviator~\f{\fsigma'}, von Mises equivalent stress~\f{\sigmavM=\sqrt{3/2}\, \|\fsigma'\|} and McAuley brackets \f{\langle \bullet \rangle = (\bullet+|\bullet|)/2}. Furthermore, \f{\dot \varepsilon_0} denotes the reference strain rate, \f{\sigma^{\rm D}} is the drag stress, \f{\sigma_{\rm y}} describes the yield stress and \f{p} designates the rate-sensitivity exponent. A rate-dependent flow rule (Eq.~\eqref{plstmdl}\f{_3}) is chosen to formally ensure differentiability of the algorithmic tangent and consequentially continuity of the Levenberg Marquardt residual \citep[see also the remark on p.~13 of ][]{wulfinghoff2026computational}. Rate-independent behavior with
\begin{equation}
    f=\sigmavM-\sigmay \leq 0, \ \ \ \lambda\geq 0, \ \ \ \lambda f =0, \ \ \ \dot \fepsp = \lambda \pd{f}{\fsigma},
\end{equation} 
where \f{f} is the yield function and~\f{\lambda} denotes the plastic multiplier can be approximated by a corresponding choice of material parameters (in this work: \f{\sigmaD=0.1}~MPa \f{\ll \sigmay}, \f{p=20}, \f{\dot \varepsilon_0=0.01}~s\f{^{-1}}). Two hardening laws are considered in this contribution. First, power law hardening is introduced as
\begin{equation}\label{eq:powerlaw}
\sigmay(\alpha) =K\left(\alpha + \alpha_0\right)^n
\end{equation}
with $K$ acting as strength coefficient, $n$ is the strain hardening exponent and $\alpha$ describes the von Mises equivalent plastic strain. Here, \f{\alpha_0} is considered for numerical regularization purposes and is set to a small value in the implementation (in this work: $\alpha_0=10^{-8}$). As a second option, Voce hardening is considered as
\begin{equation}
\sigmay(\alpha)=\sigma_{\rm y0}+(\sigma_{\infty}-\sigma_{\rm y0})\left(1-e^{-\frac{\theta_0-\theta_{\infty}}{\sigma_{\infty}-\sigma_{\rm y0}}\alpha}\right)+\theta_{\infty}\alpha,
\end{equation}
where $\sigma_{\rm y0}$ is the initial yield stress and $\sigma_{\infty}$, $\theta_0$ and $\theta_{\infty}$ are further material parameters.

\subsection{Evaluation of training and ROM parameter influence}
\subsubsection{General setup for evaluation}
\begin{figure}[h]
\centering

\begin{minipage}[t]{0.48\textwidth}
    \vspace{0pt}
    \centering
    \psfrag{C}[c][c][0.8]{a)}
    \psfrag{A}[r][c][0.8]{Al-6061-O}
    \psfrag{S}[r][c][0.8]{Steel 304}
    \psfrag{T}[r][c][0.8]{Ti-6Al-4V}
    \psfrag{V}[r][c][0.8]{Voce (mounting plate)}
    \psfrag{E}[tc][c][0.8]{$\alpha$ [-]}
    \psfrag{U}[bc][tc][0.8]{$\sigmay$ [MPa]}

    \psfrag{0}[c][l][0.7]{0}
    \psfrag{200}[c][l][0.7]{200}
    \psfrag{400}[c][l][0.7]{400}
    \psfrag{600}[c][l][0.7]{600}
    \psfrag{800}[c][l][0.7]{800}
    \psfrag{1000}[c][l][0.7]{1000}
    \psfrag{1200}[c][l][0.7]{1200}
    \psfrag{1400}[c][l][0.7]{1400}

    \psfrag{0.02}[c][bc][0.7]{0.02}
    \psfrag{0.04}[c][bc][0.7]{0.04}
    \psfrag{0.06}[c][bc][0.7]{0.06}
    \psfrag{0.08}[c][bc][0.7]{0.08}
    \psfrag{0.1}[c][bc][0.7]{0.1}
    
    \includegraphics[width=\textwidth]{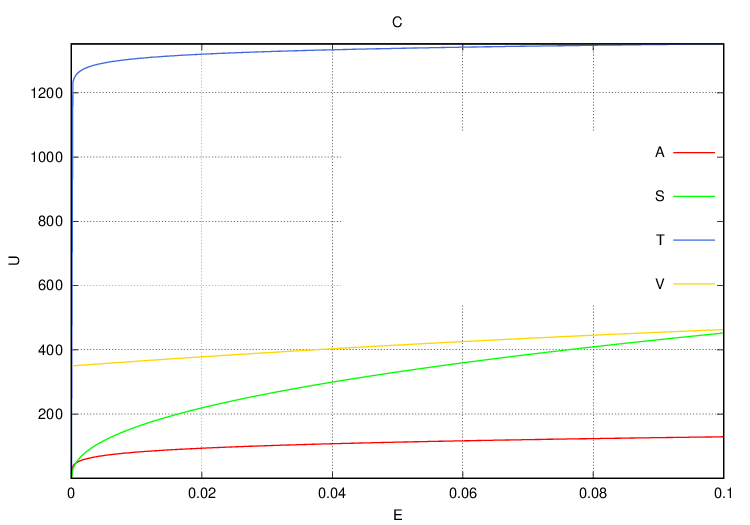}
    \label{fig:bild_a}
\end{minipage}
\hfill
\begin{minipage}[t]{0.48\textwidth}
    \vspace{0pt}
    \centering
    \psfrag{C}[c][c][0.8]{b)}
    \psfrag{E}[tc][c][0.8]{$\alpha$ [-]}
    \psfrag{U}[bc][tc][0.8]{$\sigmay$ [MPa]}

    \psfrag{0}[c][l][0.7]{0}
    \psfrag{200}[c][l][0.7]{200}
    \psfrag{400}[c][l][0.7]{400}

    \psfrag{0.02}[c][bc][0.7]{0.02}
    \psfrag{0.04}[c][bc][0.7]{0.04}
    \psfrag{0.06}[c][bc][0.7]{0.06}
    \psfrag{0.08}[c][bc][0.7]{0.08}
    \psfrag{0.1}[c][bc][0.7]{0.1}
    
    \includegraphics[width=\textwidth]{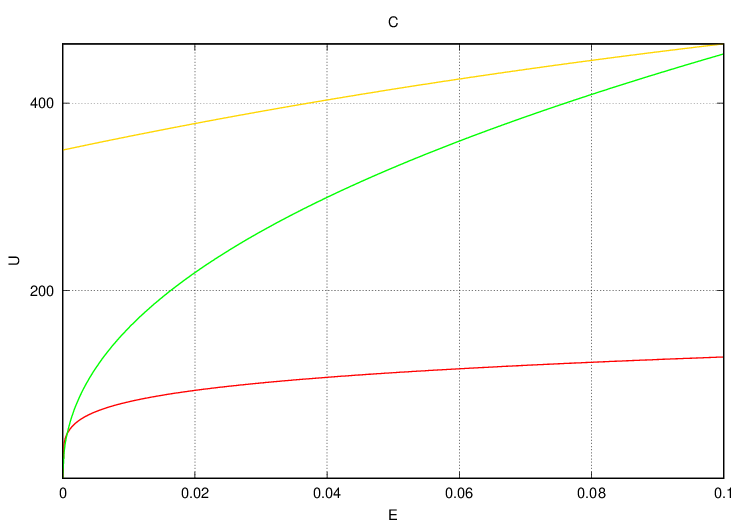}
    \label{fig:bild_b}
\end{minipage}

    \caption{a) Hardening curves for the different parameter sets in Tab.~\ref{matParam} and for the material introduced in Tab.~\ref{matParamMount}. b) Enlarged view of the smaller yield stress region.}
    \label{fig:hardeningmaterials}

\end{figure}

To investigate the influence of the training data amount and ROM parameters, a parameter study is conducted to evaluate the effect of the number of training simulations and the number of modes and generalized integration points on the accuracy for the challenging example of a porous microstructure. The mesh consists of 10061 quadratic tetrahedra with a pore volume fraction of $23.85\%$, as shown in Fig.~\ref{masse4hole} (right). Four different training sets, consisting of 22, 55, 100, and 202 microscopic FEM simulations, are compared. The LHS-generated strain paths exhibit a maximum strain norm of \f{\|\bar \feps \|_{\rm max}=0.05} for each individual path. Three different material data sets are evaluated here, with data taken from \citet{kalpakjian_manufacturing_2023} and shown in Tab~\ref{matParam}. Additionally, the power law hardening plots for the three materials are shown in Fig.~\ref{fig:hardeningmaterials}.
\begin{table}[h]
  \centering
  \small
  \begin{tabular}{||c|c|c|c|c||}
  \hline
    Material & \f{E} [GPa] & \f{\nu} & \f{K} [MPa] & \f{n}\\ 
   \hline 
   Aluminum 6061-O & 69 & 0.33 & 205 & 0.2 \\
    \hline 
   Steel 304 stainless, annealed & 193 & 0.29 & 1275 & 0.45 \\
    \hline 
   Titanium-6Al-4V, annealed, 20°C & 114 & 0.34 & 1400 & 0.015 \\
  \hline
  \end{tabular}
  \caption{Different material parameters used for porous microstructure evaluation.}
  \label{matParam}
\end{table}

The E3C results are compared with the FEM reference results using a validation data set. This data set contains 100 strain paths that have not been used during the training stage. The RMS-error is defined by
\begin{equation}\label{eq:RMSError}
e_{\rm RMS}=\sqrt{\frac{\sum_{\rm sim}\sum_{t}\normg{\bar\fsigma^{\rm FEM}-\bar\fsigma^{\rm E3C}}^{2}}{\sum_{\rm sim}\sum_{t}\normg{\bar\fsigma^{\rm FEM}}^{2}}} \cdot 100 \%.
\end{equation}

\subsubsection{Evaluation of E3C results with respect to FEM reference model}
\begin{figure}[htbp]
    \centering

    \begin{minipage}[t]{0.48\textwidth}
    \vspace{0pt}
    \centering
    
    \psfrag{X}[r][c][0.8]{22 training simulations}
    \psfrag{L}[r][c][0.8]{55 training simulations}
    \psfrag{Y}[r][c][0.8]{100 training simulations}
    \psfrag{Z}[r][c][0.8]{202 training simulations}
    \psfrag{M}[tc][c][0.8]{Modes/generalized IPs}
    \psfrag{E}[bc][c][0.8]{RMS-error in [\%]}
    \psfrag{A}[c][c][0.8]{a) Al-6061-O}
    
    \psfrag{ 10}[tc][c][0.7]{10/10}
    \psfrag{ 20}[tc][c][0.7]{20/20}
    \psfrag{ 30}[tc][c][0.7]{30/30}
    \psfrag{ 40}[tc][c][0.7]{40/40}
    \psfrag{ 50}[tc][c][0.7]{50/50}
    
    \psfrag{ 0}[r][c][0.7]{0}
    \psfrag{ 1}[r][c][0.7]{1}
    \psfrag{ 2}[r][c][0.7]{2}
    \psfrag{ 3}[r][c][0.7]{3}
    \psfrag{ 4}[r][c][0.7]{4}
    \psfrag{ 5}[r][c][0.7]{5}
    \psfrag{ 6}[r][c][0.7]{6}
    \psfrag{ 7}[r][c][0.7]{7}
    
    \includegraphics[width=\textwidth]{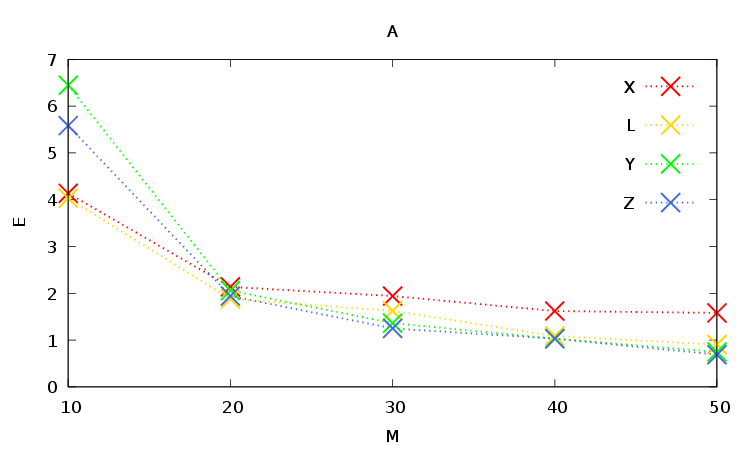}
    
    \end{minipage}
    \hfill
    \begin{minipage}[t]{0.48\textwidth}
    \vspace{0pt}
    \centering
    
    \psfrag{X}[r][c][0.8]{22 training simulations}
    \psfrag{L}[r][c][0.8]{55 training simulations}
    \psfrag{Y}[r][c][0.8]{100 training simulations}
    \psfrag{Z}[r][c][0.8]{202 training simulations}
    \psfrag{M}[tc][c][0.8]{Modes/generalized IPs}
    \psfrag{E}[bc][c][0.8]{RMS-error in [\%]}
    \psfrag{S}[c][c][0.8]{b) Steel 304}
    
    \psfrag{ 10}[tc][c][0.7]{10/10}
    \psfrag{ 20}[tc][c][0.7]{20/20}
    \psfrag{ 30}[tc][c][0.7]{30/30}
    \psfrag{ 40}[tc][c][0.7]{40/40}
    \psfrag{ 50}[tc][c][0.7]{50/50}
    
    \psfrag{ 0}[r][c][0.7]{0}
    \psfrag{ 1}[r][c][0.7]{1}
    \psfrag{ 2}[r][c][0.7]{2}
    \psfrag{ 0.}[c][c][0.7]{ }
    \psfrag{ 1.}[c][c][0.7]{ }
    \psfrag{2}[c][c][0.7]{ }
    \psfrag{4}[c][c][0.7]{ }
    \psfrag{6}[c][c][0.7]{ }
    \psfrag{8}[c][c][0.7]{ }
    
    \includegraphics[width=\textwidth]{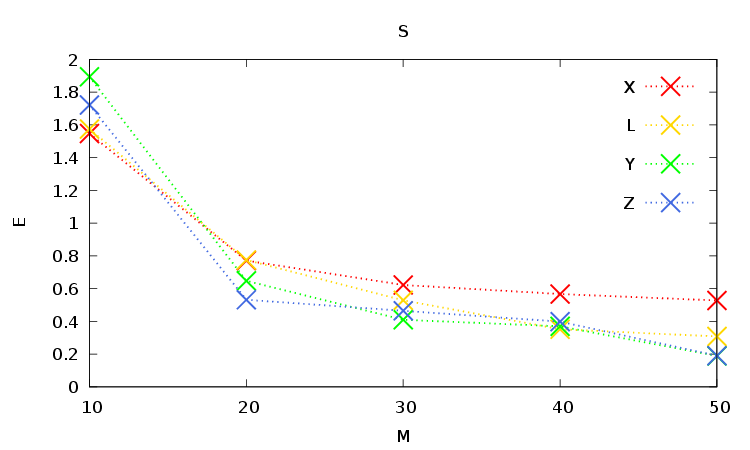}
    
    \end{minipage}
    
    \vspace{0.3cm}
    
    \begin{minipage}[t]{0.48\textwidth}
    \vspace{0pt}
    \centering
    
    \psfrag{X}[r][c][0.8]{22 training simulations}
    \psfrag{L}[r][c][0.8]{55 training simulations}
    \psfrag{Y}[r][c][0.8]{100 training simulations}
    \psfrag{Z}[r][c][0.8]{202 training simulations}
    \psfrag{M}[tc][c][0.8]{Modes/generalized IPs}
    \psfrag{E}[bc][c][0.8]{RMS-error in [\%]}
    \psfrag{T}[c][c][0.8]{c) Ti-6Al-4V}
    
    \psfrag{ 10}[tc][c][0.7]{10/10}
    \psfrag{ 20}[tc][c][0.7]{20/20}
    \psfrag{ 30}[tc][c][0.7]{30/30}
    \psfrag{ 40}[tc][c][0.7]{40/40}
    \psfrag{ 50}[tc][c][0.7]{50/50}
    
    \psfrag{ 1}[r][c][0.7]{1}
    \psfrag{ 2}[r][c][0.7]{2}
    \psfrag{ 3}[r][c][0.7]{3}
    \psfrag{ 4}[r][c][0.7]{4}
    \psfrag{ 5}[r][c][0.7]{5}
    \psfrag{ 6}[r][c][0.7]{6}
    \psfrag{ 7}[r][c][0.7]{7}
    \psfrag{ 8}[r][c][0.7]{8}
    \psfrag{ 9}[r][c][0.7]{9}
    
    \includegraphics[width=\textwidth]{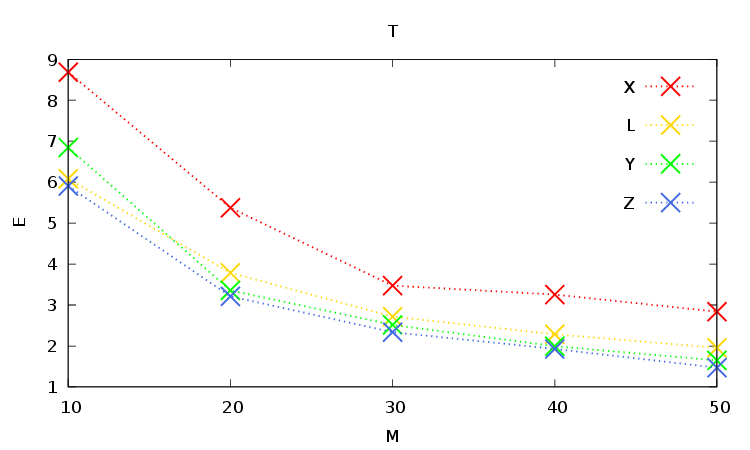}
    
    \end{minipage}
    \hfill
    \begin{minipage}[t]{0.48\textwidth}
    \vspace{0pt}
    \centering
    
    \psfrag{A}[r][c][0.8]{Al-6061-O}
    \psfrag{S}[r][c][0.8]{Steel 304}
    \psfrag{T}[r][c][0.8]{Ti-6Al-4V-20C}
    \psfrag{M}[tc][c][0.8]{Modes/generalized IPs}
    \psfrag{E}[bc][c][0.8]{RMS-error in [\%]}
    \psfrag{C}[c][c][0.8]{d) Comparison for 202 training simulations}
    
    \psfrag{ 10}[tc][c][0.7]{10/10}
    \psfrag{ 20}[tc][c][0.7]{20/20}
    \psfrag{ 30}[tc][c][0.7]{30/30}
    \psfrag{ 40}[tc][c][0.7]{40/40}
    \psfrag{ 50}[tc][c][0.7]{50/50}
    
    \psfrag{ 0}[r][c][0.7]{0}
    \psfrag{ 1}[r][c][0.7]{1}
    \psfrag{ 2}[r][c][0.7]{2}
    \psfrag{ 3}[r][c][0.7]{3}
    \psfrag{ 4}[r][c][0.7]{4}
    \psfrag{ 5}[r][c][0.7]{5}
    \psfrag{ 6}[r][c][0.7]{6}
    
    \includegraphics[width=\textwidth]{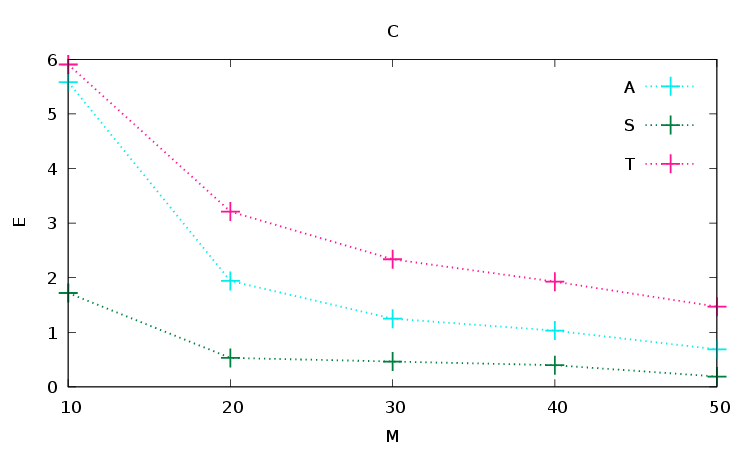}
    
    \end{minipage}
    
    \caption{a) to c) RMS-error for three different materials, highlighting the decrease in RMS-error with an increasing number of modes, generalized IPs, and training simulations. d) Comparison of the different materials for 202 training simulations, demonstrating the impact of material parameters on the accuracy of E3C results.}
    \label{fig:vierplots}
    
\end{figure}

Figure~\ref{fig:vierplots} a) to c) illustrates the influence of the number of training simulations, modes and generalized IPs on the error \eqref{eq:RMSError} for the three different materials introduced in the previous section. For simplicity, the same number of modes and generalized IPs is chosen, based on earlier experience. In general, the largest errors are seen with the smallest set of 10 modes and 10 generalized IPs, regardless of the snapshot set investigated. In addition, the set of 22 training simulations leads mostly to slightly less accurate results than snapshot sets built by more simulations. Nevertheless, in certain applications, the set of 22 training simulations might be sufficient for materials showing a strong strain hardening, such as Steel 304 (see Fig.~\ref{fig:vierplots}), with errors reaching values below \f{1\%} in this investigation. In contrast, for Ti-6Al-4V, a material with very little hardening (\f{n=0.015}) the 22 training simulations result in the largest errors seen for this material of \f{\sim\!3\%} for 50 modes/IPs, while a reduction is achieved by using 55 training simulations with errors of \f{\sim\!2\%}. The sets of 100 and 202 training simulations show some minor difference in errors, the gain being small compared to the jump from 22 to 55 simulations. In case of Al-6061-O, the same trends are found. Generally, an increasing number of modes and generalized IPs leads to a reduction of the RMS-error. This trend is evident for all materials and is shown in Fig.~\ref{fig:vierplots} d), which compares the different materials for the set of 202 training simulations. In general, it can be concluded, that for the materials and combination of modes and generalized IPs investigated here, a training dataset from $50$ simulations represents a good compromise between accuracy and offline effort.

\subsection{Porous mounting plate}
\subsubsection{Macroscopic problem setup}
\begin{figure}[h]
\centering
  \includegraphics[width=0.9\textwidth]{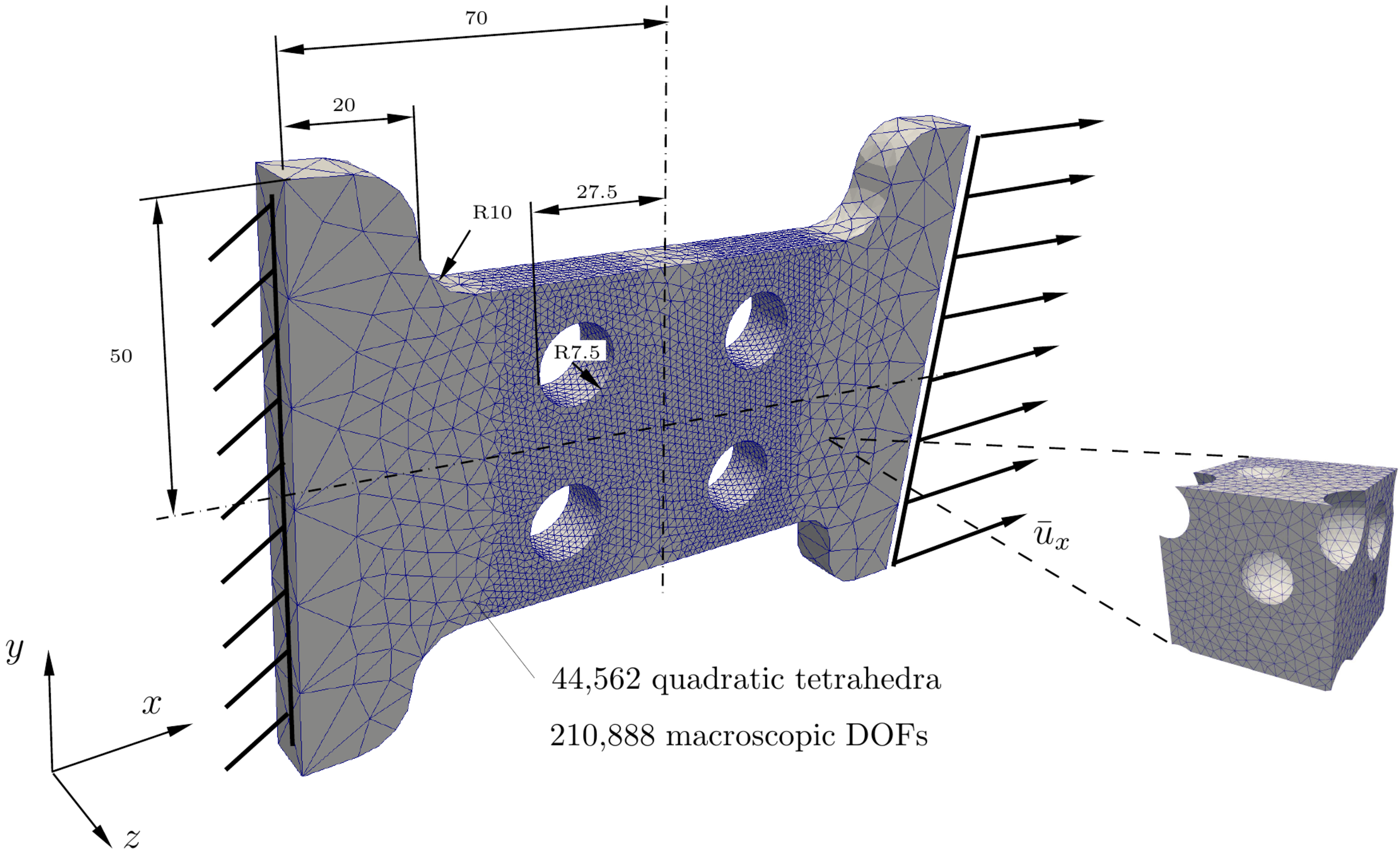}
	\caption{Dimensions, mesh and boundary conditions of the porous mounting plate.}
\label{masse4hole}
\end{figure}

In this section, a tensile test of the porous mounting plate shown in Fig.~\ref{masse4hole} is simulated, assuming a maximum strain of 1\% and free lateral contractions. Quadratic tetrahedra are used on both scales. The study compares the performance of two element types on the macroscale. The first element type uses four integration points (standard integration) in all 44,562 elements, leading to a total number of 4\f{\times}44,562=178,248 microscopic problems to be solved. The second element type is a stabilized mean strain element (for details, see \ref{appmeanstrain}), which has the advantage that the microscopic problem needs to be solved only once per element. 
\subsubsection{Microscopic problem setup}
On the microscale, the viscoelastoplastic material model with Voce hardening, described in Sect.~\ref{sectConst}, is applied. The material parameters are summarized in Tab.~\ref{matParamMount}.
\begin{table}[h]
  \centering
  \small
  \begin{tabular}{||c|c|c|c|c|c||}
  \hline
   \f{E} [GPa] & \f{\nu} & \f{\sigma_{\rm y 0}} [MPa] & \f{\sigma_{\rm y \infty}} & \f{\theta_0} [MPa] & \f{\theta_{\infty}} [MPa]\\ 
   \hline 
   210 & 0.3 & 349.9 &  599.9 & 1500 & 10 \\
  \hline
  \end{tabular}
  \caption{Material parameters for mounting plate example}
  \label{matParamMount}
\end{table}
The microscopic mesh, shown in Fig.~\ref{masse4hole} (right), consists of 10,061 quadratic tetrahedra with 44,112 degrees of freedom. The pore volume fraction is 23.85~\%.
The snapshot data has been generated using the Latin Hypercube Sampling procedure described in Sect.~\ref{Sect:Snapshots} with a maximum strain norm of \f{\|\bar \feps \|_{\rm max}=0.05}. Two training sets are investigated, comprising 210 and 49 microscopic FEM simulations, respectively. The E3C model is trained using 50 modes and 50 generalized IPs. This non-adaptive model is compared with its adaptive counterpart with Sequential Subspace Mode Adaptation (see Sect.~\ref{sec:SSMA_1}) with five base modes and four adaptive modes (while retaining the 50 generalized IPs). In other words, the first five modes are identical to the first five modes of the 50-mode non-adaptive model. The four adaptive modes are identified online out of the space spanned by the remaining 45 modes by the procedure described in Sect.~\ref{sectadaptmodes}. To limit the computational overhead due to the mode adaptation, the first two macroscopic Newton iterations are performed using the five base and four adaptive modes identified in the previous time step. The adaptive modes are then updated during the third macroscopic Newton iteration and all remaining Newton iterations are performed with the new fixed set of five base and four adaptive modes of the current time step. 
\subsubsection{Hard- and Software}
The CPU used throughout this section was an Intel$^{\textregistered}$ Core\f{^{\rm TM}} i9-13980HX CPU with 64 GB RAM (laptop hardware). All simulations were performed with an in-house FEM-code. The microscopic FEM problems for the snapshot collection were solved in parallel using OpenMP \citep{dagum1998openmp}. Each of the 19 threads solved the related linear equation systems via a serial Jacobi-preconditioned CG solver. The E3C modes were computed using an in-house Levenberg-Marquardt implementation. The two-scale simulation used 17 threads for the macroscopic assembly, involving the solution of the microscopic problems and the PARDISO solver \citep{schenk2006fast,intel_mkl} for the solution of the macroscopic linear equation systems. Additionally paraview \citep{ahrens2005paraview} and gnuplot \citep{gnuplot54} have been used for visualization of the corresponding results.
The standalone code performing FEM-simulations for snapshot collection, singular value decomposition, clustering, Levenberg-Marquardt optimization as well as the two-scale simulation, can be downloaded from a git-repository \citep{CodeE3CRef_Plasti}. All simulation data is available upon reasonable request.

\subsubsection{Simulation results}
\begin{figure}[h]
\centering
	\hspace{-10mm}
  \psfrag{40360}{1)}
  \psfrag{32439}{2)}
  \psfrag{41068}{3)}
  \psfrag{41453}{4)}
  \psfrag{13048}{5)}
  \psfrag{40560}{6)}
  \psfrag{SVM}{\footnotesize \f{\sigmavM} [MPa]}
  \psfrag{0.9}{\footnotesize 0.9}
  \psfrag{301.1}{\footnotesize 301.1}
  \psfrag{alpha}{\footnotesize \f{\alpha}}
  \psfrag{0.0}{\footnotesize 0.0}
  \psfrag{0.68}{\footnotesize 0.68}
  \psfrag{}{}
\ifthenelse{\boolean{badQuality}}{
  \includegraphics[width=0.7\textwidth]{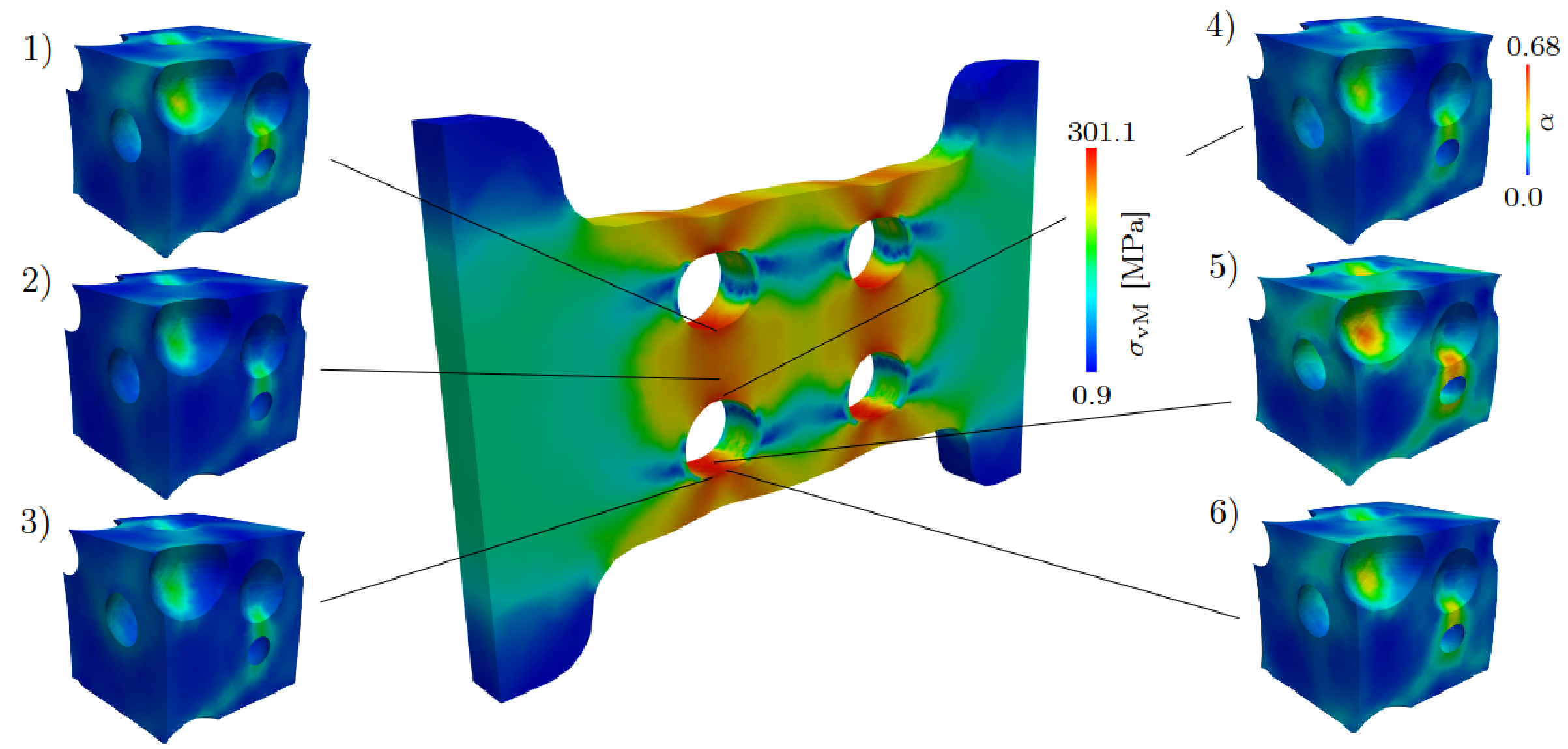}
}{
  \includegraphics[width=0.9\textwidth]{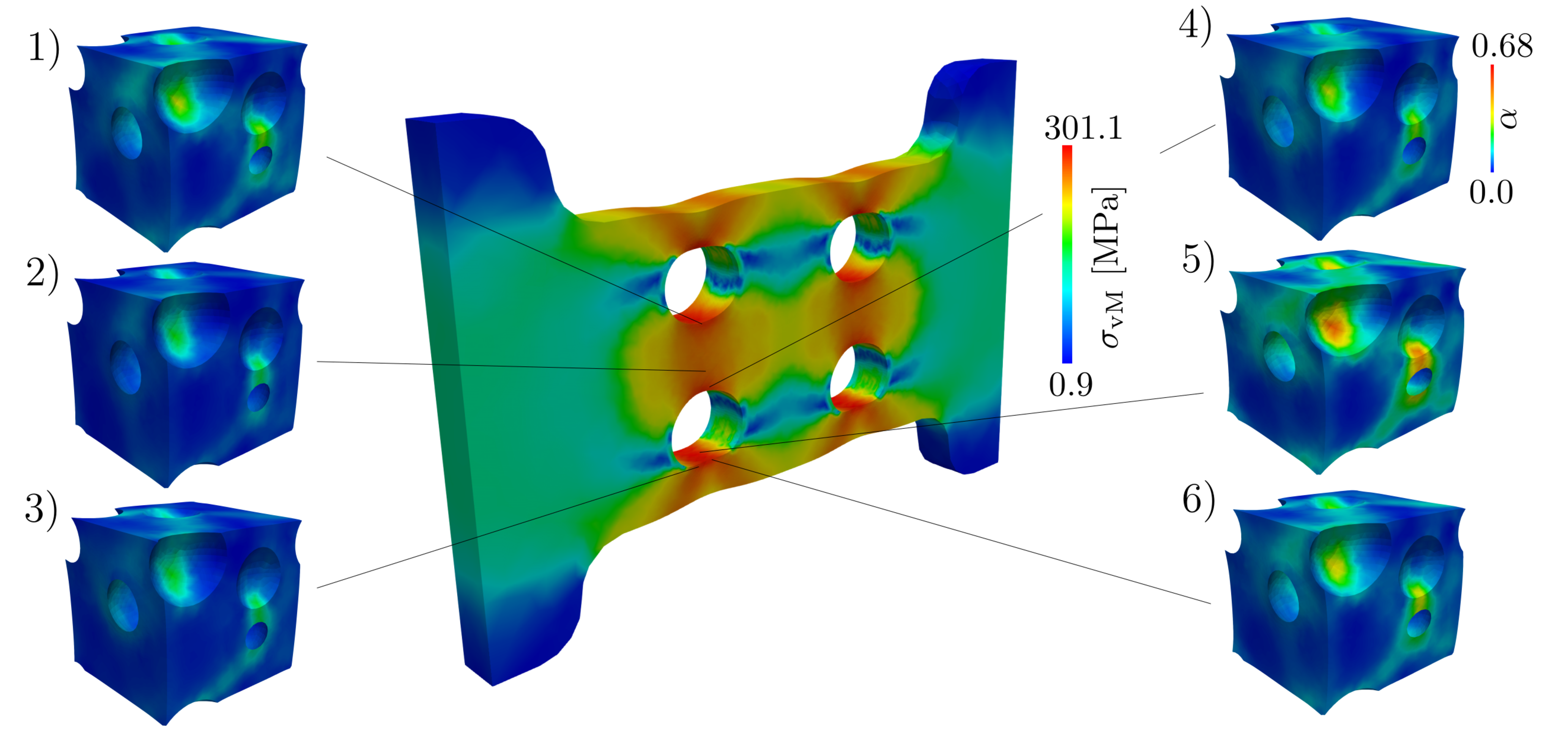}
}
\caption{Two-scale simulation results with a macroscopic deformation scale factor of 5 (the microscopic deformations are unscaled).}
\label{mountingPlatePoresOverview}
\end{figure}

\begin{figure}[h]
\centering
\includegraphics[width=0.9\textwidth]{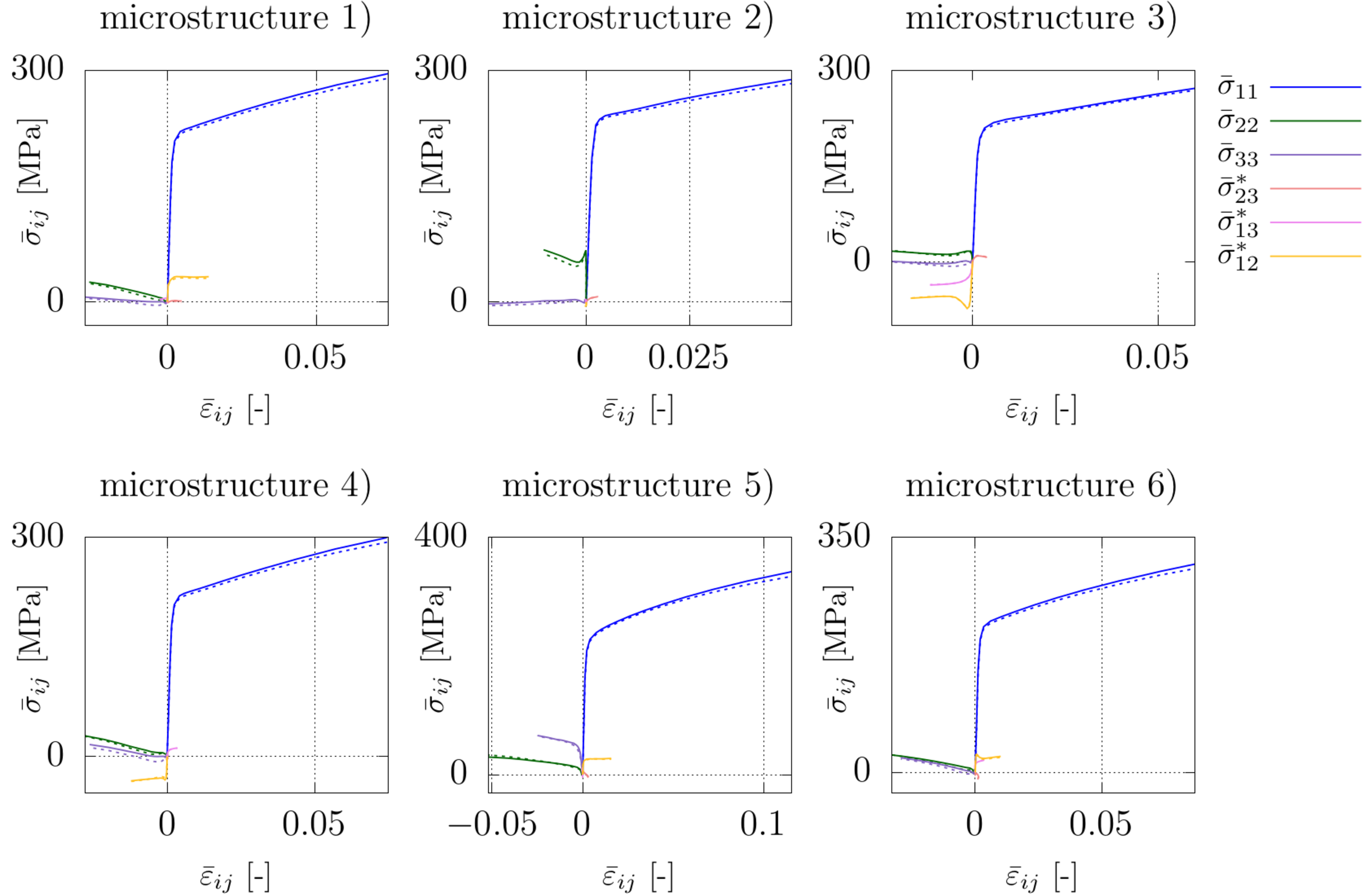}
	\caption{Comparison of E3C predictions (full lines) with FEM (dashed lines) for microstructures 1)-6) in Fig.~\ref{mountingPlatePoresOverview}. The shear stresses/strains are given in Mandel notation (\f{\bar\sigma^*_{ij}=\sqrt{2}/2\bar\sigma_{ij}}).}
\label{strssstrncrvs}
\end{figure}

The overall deformation and macroscopic von Mises stress at the end of the simulation are depicted in Fig.~\ref{mountingPlatePoresOverview} (center). On the microscale, the non-adaptive E3C model with 50 modes trained on 210 simulations for snapshot collection was applied. The figure also presents the microscopic deformations and equivalent plastic strain distributions at six macroscopic positions\footnote{The depicted microscopic contour plots are FEM results obtained by postprocessing the macroscopic strain data collected during the two-scale simulation.}. The results show that the small strain hypothesis is violated on the microscale. Nevertheless, they remain of interest, as they demonstrate the robustness of the two-scale simulation under comparatively severe deformation conditions.
The macroscopic stress strain curves of the six microscopic problems shown in Fig.~\ref{mountingPlatePoresOverview} are presented in Fig.~\ref{strssstrncrvs}. The E3C predictions (solid lines) are compared with the corresponding FEM results (dashed lines), indicating a good agreement\footnote{The FEM model requires smaller time steps than the E3C model. For that reason, the E3C results were also obtained in a postprocessing step, and the time steps were decreased correspondingly, in order to rule out any influence of the time discretization.}. The diagrams reveal that the norm of the macroscopic strain succeeds the maximum value of 0.05 applied during offline training stage, in some cases by more than 100~\%. The agreement with the FEM results is hardly affected by that circumstance, suggesting good extrapolation properties of the E3C hyper-reduced model.

\subsubsection{Online and offline performance}
\begin{figure}[h]
\centering
  \includegraphics[width=0.9\textwidth]{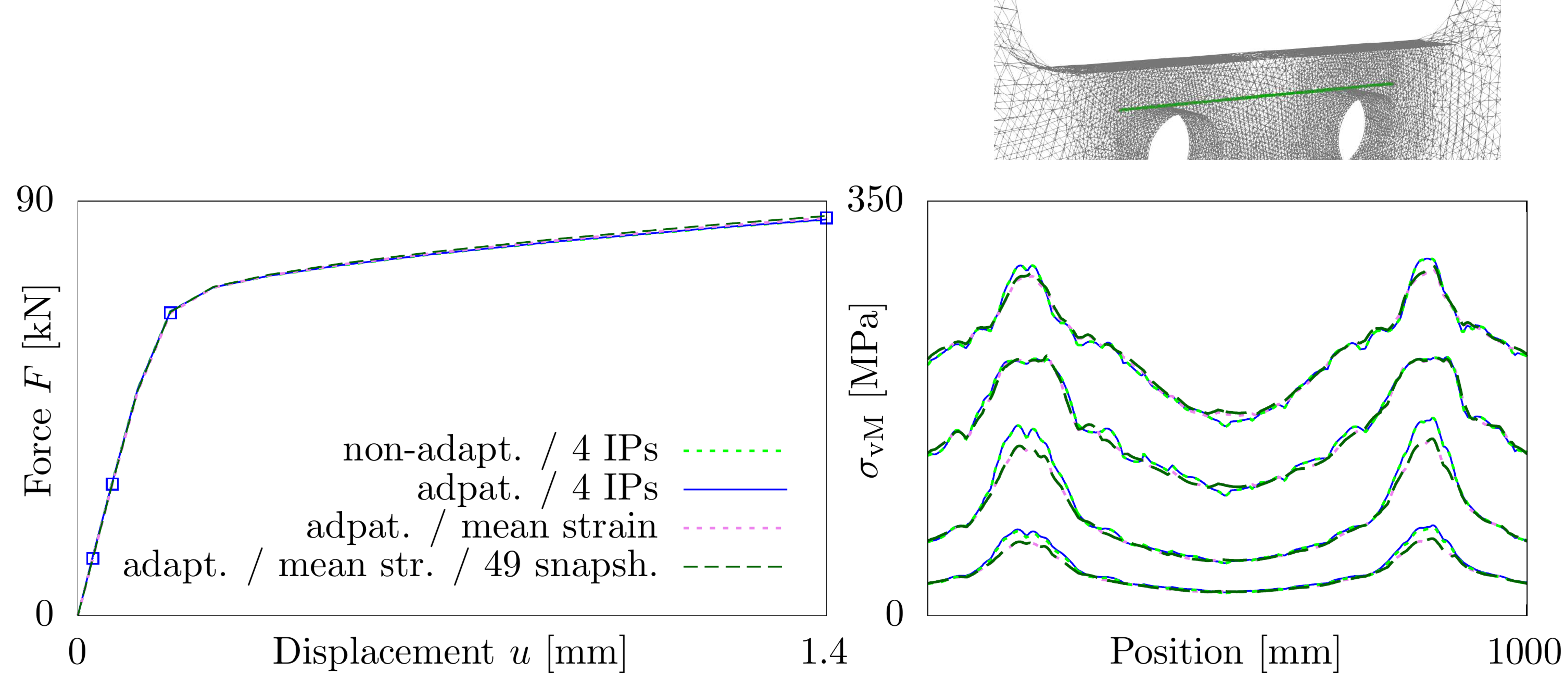}
	\caption{Left: comparison of the reaction force for the different formulations. Right: von Mises equivalent stress plotted along the green line (upper right picture) for the four states illustrated by the squares in the left diagram.}
\label{reactnfrcs}
\end{figure}

\begin{figure}[h]
\centering

\begin{minipage}[t]{0.48\textwidth}
    \vspace{0pt}
    \centering
    \psfrag{A}[cl][c][0.8]{4 IPs/non-adaptive}
    \psfrag{B}[cl][c][0.8]{4 IPs/adaptive}
    \psfrag{C}[cl][c][0.8]{mean strain/adaptive}
    \psfrag{D}[cl][c][0.8]{single scale (estim.)}
    \psfrag{E}[c][c][0.8]{a) Online effort}
    \psfrag{F}[bc][tc][0.8]{wall clock time [min]}

    \psfrag{0}[c][cl][0.7]{0}
    \psfrag{10}[c][cl][0.7]{10}
    \psfrag{20}[c][cl][0.7]{20}
    \psfrag{30}[c][cl][0.7]{30}
    \psfrag{40}[c][cl][0.7]{40}
    \psfrag{50}[c][cl][0.7]{50}
    \psfrag{60}[c][cl][0.7]{60}
    \psfrag{70}[c][cl][0.7]{70}
    \psfrag{80}[c][cl][0.7]{80}

    \includegraphics[width=\textwidth]{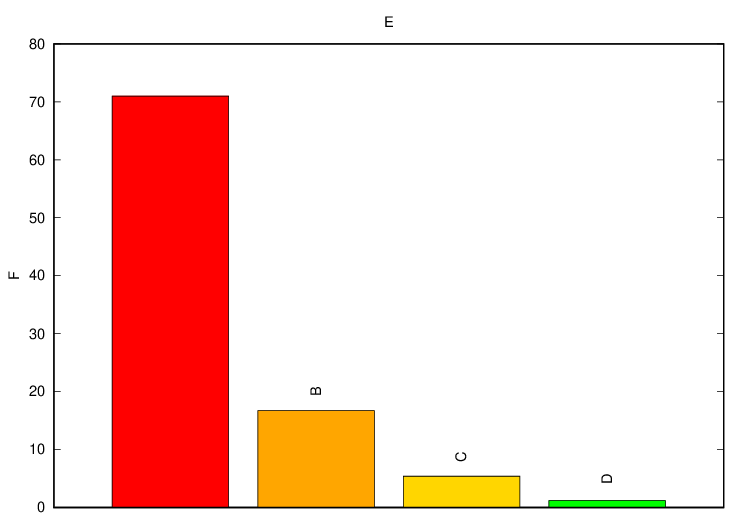}
\end{minipage}
\hfill
\begin{minipage}[t]{0.48\textwidth}
    \vspace{0pt}
    \centering
    \psfrag{A}[cl][c][0.8]{210 snapsh. sim.}
    \psfrag{B}[cl][c][0.8]{49 snapsh. sim.}
    \psfrag{E}[c][c][0.8]{b) Snapshot collection/ LM}
    \psfrag{F}[bc][ct][0.8]{wall clock time [min]}

    \psfrag{0}[c][cl][0.7]{0}
    \psfrag{20}[c][cl][0.7]{20}
    \psfrag{40}[c][cl][0.7]{40}
    \psfrag{60}[c][cl][0.7]{60}
    \psfrag{80}[c][cl][0.7]{80}
    \psfrag{100}[c][cl][0.7]{100}
    \psfrag{120}[c][cl][0.7]{120}
    \psfrag{140}[c][cl][0.7]{140}
    \psfrag{160}[c][cl][0.7]{160}
    \psfrag{180}[c][cl][0.7]{180}
    
    \includegraphics[width=\textwidth]{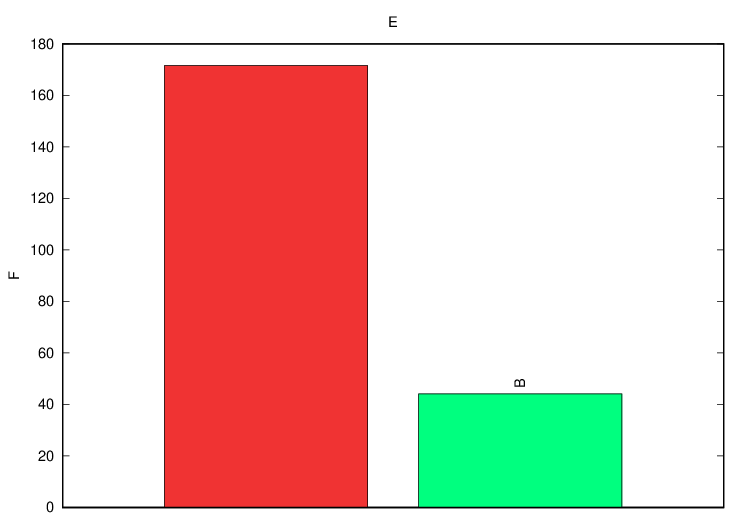}
\end{minipage}

    \caption{a) Online effort, as described by the Tab.~\ref{porestimes}. Starting with 4 IPs at the macroscale and the conventional E3C hyper-reduction on the microscale (red column) the online effort can be decreased with 'Sequential Subspace Mode Adaptation' (orange column). Additional application of the mean strain element (yellow column) leads to online speeds even closer to single scale(green column). b) Exemplary offline effort for 210 and 49 snapshot simulations.}
    \label{fig:comptimes}

\end{figure}

In order to investigate the on- and offline performance, the non-adaptive E3C model with 50 modes is compared to its adaptive counterpart with five base and four adaptive modes. Initially, four macroscopic integration points (IPs) are applied per element. The macroscopic force-displacement diagrams are depicted in Fig.~\ref{reactnfrcs} (left) in light green and blue. They are hardly distinguishable, illustrating an excellent match between non-adaptive and adaptive model. Local stress data is also provided in the right part of the figure. Specifically, the von Mises stress is plotted along the green line depicted in the upper right picture for the four states indicated by small squares in the left diagram. Also in the right diagram, an excellent agreement between both models is found, i.e., the model with Sequential Subspace Mode Adaptation matches the results of the non-adaptive model very well.
\\
The computational effort is compared in Fig.~\ref{fig:comptimes} (see also Tab.~\ref{porestimes}). The offline effort is identical for both models and is dominated by the 210 FEM simulations used for the snapshot collection, which took \f{\sim}2.5~h. However, the online effort is decreased by a factor of \f{\sim}4 from 71\unit{min} to 16.7\unit{min}, when using Sequential Subspace Mode Adaptation. It is essentially composed of contributions from the macroscopic linear solver (\f{\sim}1.2~min) and the assembly process (15.3~min). If one assumes that the computational effort of a purely macroscopic simulation, using a phenomenological model instead of E3C, mainly results from the linear equation solver (\f{\sim}1.2~min), then the computational effort of the E3C model with Sequential Subspace Mode Adaptation (16.7~min) is still approximately fourteen times that of a single scale simulation, as shown in Fig.~\ref{fig:comptimes}~a).
\begin{table}[h]
  \centering
  \footnotesize
  \begin{tabular}{||c||c|c|c|c||c||}
  \hline
   \parbox{2.3cm}{\centering \scriptsize   wall clock\\times [min]}& \parbox{2.3cm}{\centering \scriptsize 4 IPs/non-adapt. \\210 snapsh.~sim.} & \parbox{2.3cm}{\centering \scriptsize 4 IPs/adapt. \\ 210 snapsh.~sim.} & \parbox{2.3cm}{\centering \scriptsize mean str./adapt. \\ 210 snapsh.~sim.} & \parbox{2.3cm}{\centering \scriptsize mean str./adapt. \\ 49 snapsh.~sim.} & \parbox{2.0cm}{\centering \scriptsize \centering particle-\\reinforced} \\
   \hline \hline
   \parbox{2.3cm}{\centering \scriptsize  online (lin.~sol./\\assembly)} &71.0 (1.2/69.7) & 16.7 (1.2/15.3) &  5.4 (1.2/4.0) & -- & 3.2 (1.3/1.7) \\
  \hline
  \parbox{2.5cm}{\centering \scriptsize  offline \\ (snapshots/Lev.~M.)} & 148.1 / 23.4 & 148.1 / 23.4 & 148.1 / 23.4 & 34.6 / 9.5 & 5.8 / 0.5\\
  \hline
  \end{tabular}
  \caption{Online and offline efforts (wall clock times): times spent by linear equation solver, assembly process, snapshot collection and Levenberg-Marquardt based cost function minimization. Other offline costs are negligible and not shown. The rightmost column refers to the particle-reinforced microstructure (Sect.~\ref{sectpart}). '4~IPs': 4 macroscopic integration points, 'mean str.': mean strain element, 'adapt./non-adapt.': with/without adaptive modes, 'Lev.~M.': Levenberg-Marquardt.}
  \label{porestimes}
\end{table}
\subsubsection{Efficiency improvement through mean strain elements and reduced training}
A natural strategy for further efficiency improvement is to use reduced integration or mean strain elements, evaluating the E3C model only once (instead of four times) per macroscopic element. Details on the mean strain formulation used here are given in \ref{appmeanstrain}. The corresponding force displacement curve (dashed pink line in Fig.~\ref{reactnfrcs}, left) shows very good agreement with the fully integrated elements. The right graph also demonstrates an overall good agreement, but also reveals deviations in region of maximum stress, close to the holes of the mounting plate. While these deviations are clearly perceptible during the first two (and mostly elastic) time steps, they are less pronounced in the plastic regime (upper two curves in Fig.~\ref{reactnfrcs}, right). The stabilization stiffness (see \ref{appmeanstrain} for details) is taken constant and significantly smaller in magnitude than the elastic properties, which might explain the observed deviations in the elastic regime. More advanced stabilization techniques might lead to a better agreement.\\
The computational saving during the assembly process indeed corresponds to the already anticipated factor of \f{\sim}4, using the mean strain element, as can be seen from Tab.~\ref{porestimes}. In fact, the microscopic E3C simulations during the macroscopic assembly process take 4~min, while the linear equation solver needs 1.2~min. In other words, the two-scale simulation can be expected to take approximately four times as long as the single scale simulation for the given example, as indicated in Fig.~\ref{fig:comptimes} a). This result illustrates that two-scale simulations can significantly benefit from reduced integration schemes. However, the slight deviations observed in the linear regime call for more accurate stabilization techniques or full integration in regions of primary interest in the future.\\
Finally, the snapshot collection dominates the offline effort and shall therefore be investigated in the following. As suggested in the previous evaluation of snapshot data in Sect.~\ref{Sect:Snapshots}, \f{\thicksim}50 simulations should be sufficient for snapshot collection. Indeed, using 49 FEM simulations for snapshot collection hardly changes the results, as both the force displacement data and the local von Mises stress data almost coincide with the results obtained after training on snapshots from 210 FEM simulations. As expected, this has almost no effect on the online performance, but significantly reduces the offline effort from \f{\sim}2.86~h to 44.1~min (compare with Fig.~\ref{fig:comptimes} b)).
\subsubsection{Impact of mode adaptation on macroscopic convergence}\label{sec:Results_ConvergencEval}
Table~\ref{resnormsmacr} (left column) shows representative values of the relative L2-norm of the residual associated with the macroscopic Newton iterations at different time steps. The adaptive modes are identified during the third iteration, which is marked in blue in the table. Strictly speaking, updating the modes during this iteration implies a temporary loss of consistency of the linearization on the macroscopic level, which could lead to a perturbation in the residual. However, as the table shows, this effect is negligible and the residual continues to decrease in the subsequent iterations. This is different during the first time steps, which are also depicted, illustrating that this effect already diminishes after the first two time steps. The residual perturbation in time step 1 (second column) is larger than in the other steps, which may result from the fact that the first two Newton iterations of the very first step are carried out with the four adaptive modes being taken as modes 6 to 9 of the full 50-mode model. Nevertheless, this effect diminishes quickly and for the following steps the convergence behavior is not affected. Summarizing, the results show, that the Sequential Subspace Mode Adaptation does not noticeably disrupt the convergence.
\begin{table}[h]
  \centering
  \small
  \begin{tabular}{c||c|c|c}
  step 8 & step 1 & step 2 & step 3 \\
  \hline
1.000E+00 & 1.000E+00 & 1.000E+00 & 1.000E+00 \\
0.128E-02 & 0.765E-10 & 0.378E-04 & 0.116E-02 \\
{\cblue 0.549E-03} & {\cblue 0.748E-03} & {\cblue 0.665E-04} & {\cblue 0.543E-03} \\
0.538E-04 & 0.109E-14 & 0.270E-07 & 0.249E-04 \\
0.162E-05 &           & 0.720E-12 & 0.124E-06 \\
0.342E-08 &           &           & 0.341E-11 \\
0.124E-12 &           &           &
  \end{tabular}
  \caption{Relative residual norms of macroscopic Newton iterations for different time steps for the model with Sequential Subspace Mode Adaptation. The blue residuals mark the Newton iteration used for the adaptation of the modes.}
  \label{resnormsmacr}
\end{table}
\subsubsection{Particle-reinforced microstructure}\label{sectpart}
\begin{figure}[h]
\centering

\begin{minipage}[t]{0.26\textwidth}
    \vspace{0pt}
    \centering

    \includegraphics[width=\linewidth]{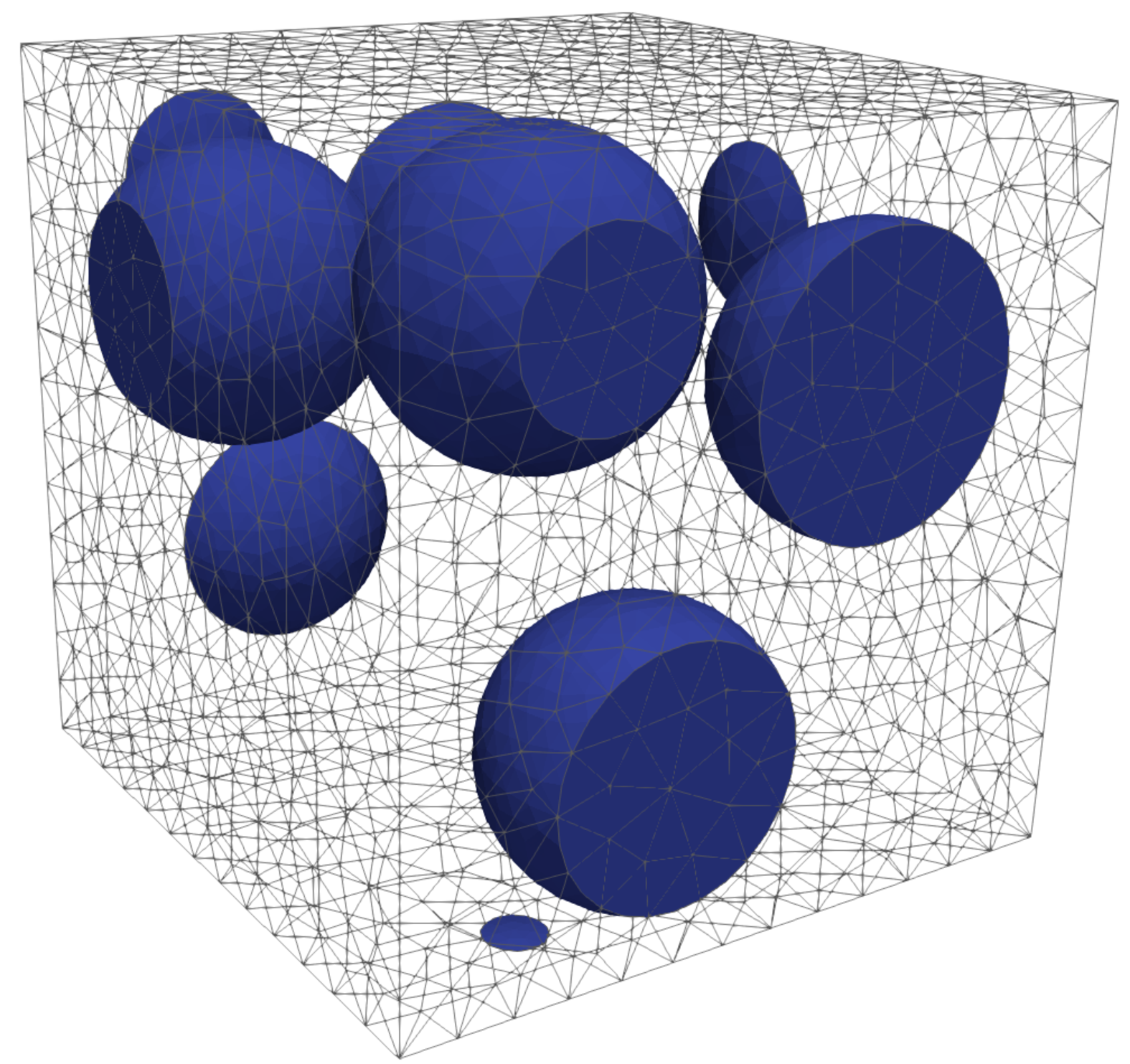}

    \vspace{0.5cm}

    \psfrag{A}[cl][c][0.7]{two-scale}
    \psfrag{B}[cl][c][0.7]{single scale}
    \psfrag{C}[cl][c][0.7]{(estim.)}
    \psfrag{E}[c][c][0.7]{}
    \psfrag{F}[bc][ct][0.7]{wall clock time [min]}

    \psfrag{0}[c][cl][0.7]{0}
    \psfrag{1}[c][cl][0.7]{1}
    \psfrag{2}[c][cl][0.7]{2}
    \psfrag{3}[c][cl][0.7]{3}
    \psfrag{4}[c][cl][0.7]{4}
    \psfrag{5}[c][cl][0.7]{5}

    \includegraphics[width=\linewidth]{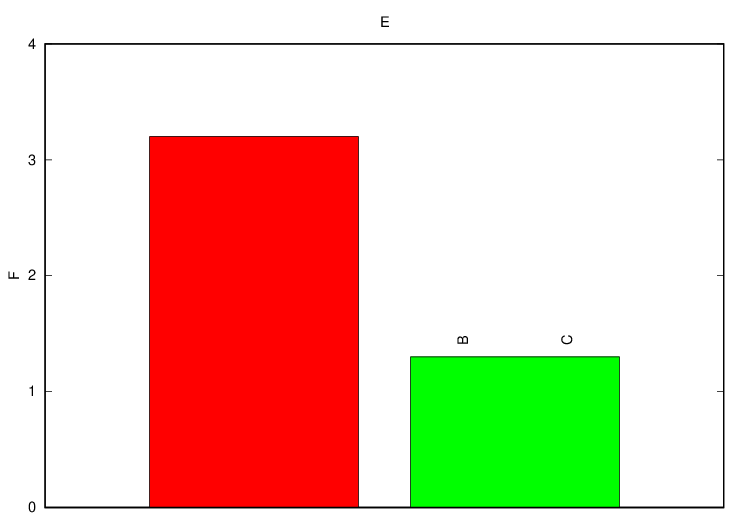}

\end{minipage}
\hfill
\begin{minipage}[t]{0.73\textwidth}
    \vspace{0pt}
    \centering

    \includegraphics[width=\linewidth]{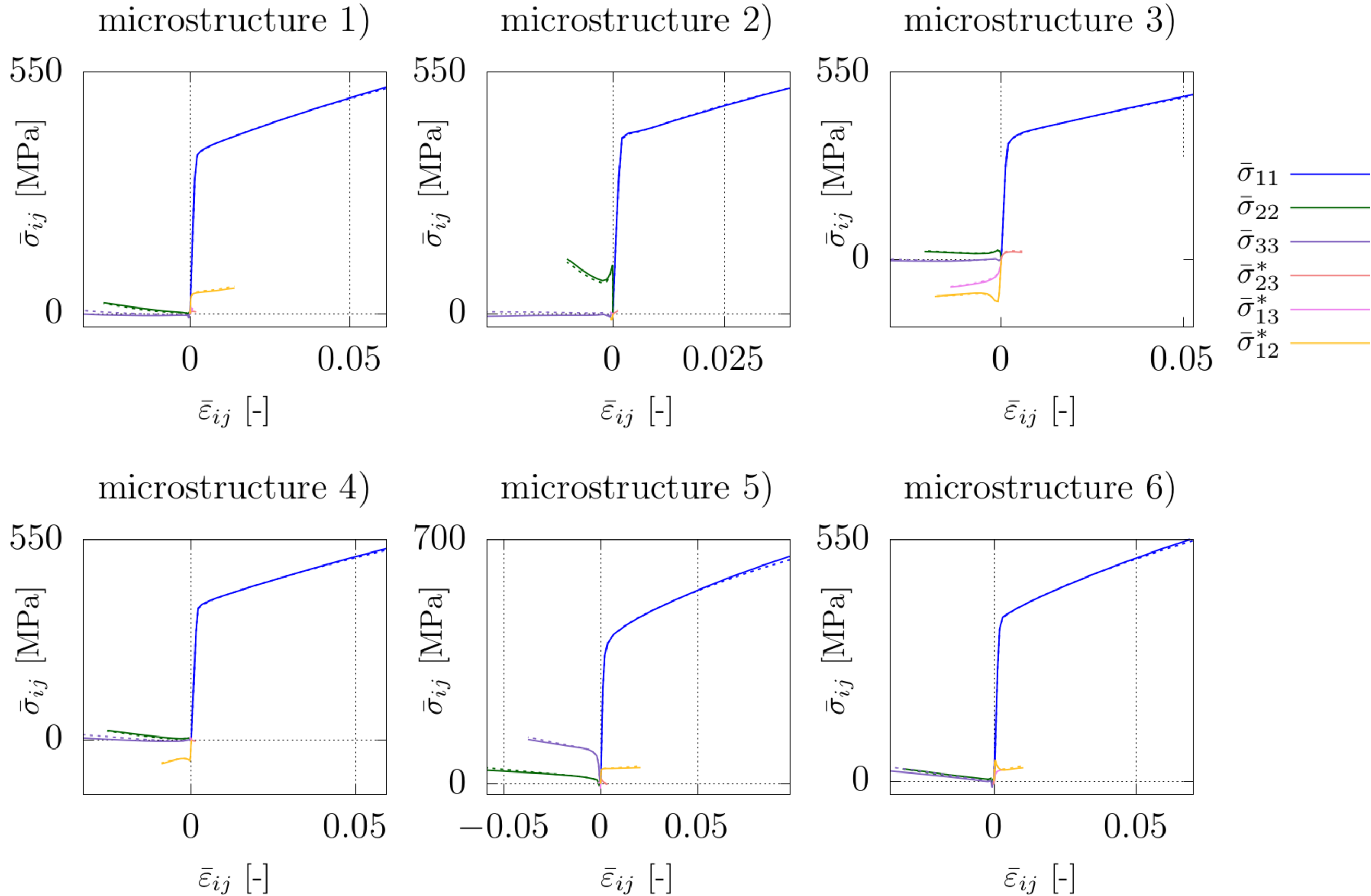}

\end{minipage}

\caption{Comparison of E3C predictions (full lines) with FEM (dashed lines) for particle-reinforced microstructures 1)-6). The shear stresses/strains are given in Mandel notation (\f{\bar\sigma^*_{ij}=\sqrt{2}/2\bar\sigma_{ij}}).}
\label{strssstrncrvs2}

\end{figure}

Next, a particle-reinforced microstructure is considered with matrix material as in the previous section and isotropic linear-elastic particles with \f{E=300}~GPa and \f{\nu=0.25}. The training was performed using 18 snapshot simulations with reduced volumetric strain (compare Sect.~\ref{Sect:Snapshots}) and a total offline simulation time of 5.8~min (rightmost column in Tab.~\ref{porestimes}). In comparison to the porous microstructure, a smaller number of 25 modes and 25 E3C integration points were identified by the Levenberg-Marquardt algorithm in 0.5~min. Combining mean strain elements with Sequential Subspace Mode Adaptation (5 base modes and 4 adaptive modes) led to an online effort of 1.7~min for the solution of the microscopic problems, while the macroscopic linear equation solver required 1.3~min, which is in line with the times observed for the porous microstructure. The local stress-strain response was evaluated for the same six positions as for the porous microstucture (Fig.~\ref{mountingPlatePoresOverview}) and shows very good agreement with corresponding FEM-simulations (see Fig.~\ref{strssstrncrvs2}).

\section{Conclusion}\label{sec:Conclusion}
Throughout this contribution, the E3C hyper-reduction is successfully applied to dissipative materials with internal variables, as demonstrated for an exemplary viscoelastoplastic material model. The key idea is to determine the E3C modes that satisfy a hyper-reduced and mode restricted version of the Hill-Mandel condition, ensuring accurate prediction of macroscopic stresses and satisfaction of the microscale equilibrium. The E3C model is constructed to yield results equivalent to those of a high-fidelity model (in this contribution a FEM model). Further, the Sequential Subspace Mode Adaptation procedure is introduced, which can be rather easily integrated into pROM-codes using linear subspace. This approach enables two-scale simulations of engineering structures, like the ones shown in the Sect.~\ref{sec:Results}, with computational efforts close to that of single scale simulations.
The Sequential Subspace Mode Adaptation defines an affine subspace during the online phase, based on the already existing linear subspace determined during the offline stage. This affine subspace is constructed based on the leading modes of the linear subspace determined offline, enriched by an online extension based on the potential gradient. In this way, deformation patterns not captured by the leading offline modes are incorporated efficiently. By using the Sequential Subspace Mode Adaptation those missing deformation patterns are included in the affine subspace. Keeping the dimension of the subspace significantly smaller than that of a conventional offline-generated subspace. In contrast to nonlinear manifold-based methods, this does not lead to a further reduction of integration points. Further research is necessary to evaluate how the proposed method would perform compared to nonlinear manifold-based methods. The benefits of this approach are demonstrated in the Sect.~\ref{sec:Results}, where two-scale simulations achieve runtimes close to the single scale case. Additionally, the parameter studies shown contain an investigation of the training data influence, highlight two different hardening behaviors to work within the framework, and show results for two representative microstructures, one porous and one containing inclusions.

\section{Acknowledgement}
This work was funded by the German Research Foundation (Deutsche Forschungsgemeinschaft, DFG) as part of “Statistically Compatible Hyper-Reduction for Multiphysics Homogenization” under Grant No. 550055706. This support is gratefully acknowledged.

\section{AI Declaration Statement}
The line of arguments in \ref{appweakcomp} has been found with the help of ChatGPT \citep{chatgpt2026}.

\section{Appendix}\label{sec:Appendix}

\begin{appendices}
  \renewcommand{\thesection}{Appendix \Alph{section}}
\section{Weak compatibility condition}\label{appweakcomp}
First, it is noted that one can show that Eq.~\eqref{weakcmpt} is indeed valid for compatible strain fields in analogy to the weak form of the equilibrium condition (Eq.~\eqref{microlmb}). It remains to explain why Eq.~\eqref{weakcmpt} is violated for an incompatible symmetric second order tensor field~\f{\feps^*}. In fact, it is sufficient to restrict the calculation to the fluctuation part~\f{\tilde\feps^*=\feps^*-\avg{\feps^*}} and to search for that compatible field~\f{\tilde\feps=\fsym{\grad{\tilde \fu}}}, which is closest to~\f{\tilde\feps^*} in the sense that the periodic field~\f{\tilde\fu} solves
\begin{equation}
    \infi{\tilde {\blds u}\#} \int\limits_\Omega \frac{1}{2} (\tilde\feps-\tilde\feps^*):\ffC:(\tilde\feps-\tilde\feps^*) \d \Omega >0 \label{enrgy}
\end{equation}
for some positive-definite "stiffness-tensor"~\f{\ffC}. In this setting, \f{\tilde \feps^*} can be interpreted as eigenstrain field and expression~\eqref{enrgy} as strain energy due to~\f{\tilde\feps^*}. At the minimum, the field~\f{\tilde\fsigma=\ffC:(\tilde\feps-\tilde\feps^*)} satisfies the conditions \f{\avg{\tilde \fsigma}=\fzero}, \f{\div{\tilde\fsigma}=\fzero} in~\f{\Omega} and~\f{[\![\tilde \fsigma]\!]\fn=\fzero} on~\f{\partial \Omega}. As a consequence of the inequality in Eq.~\eqref{enrgy}, it follows that
\begin{equation}
    \int\limits_\Omega \tilde \fsigma : (\tilde \feps-\tilde \feps^*)\d \Omega = -\int\limits_\Omega \tilde \fsigma : \tilde \feps^*\d \Omega >0, \label{ineq}
\end{equation}
where use has been made of the equilibrium character of~\f{\tilde \fsigma}. The inequality in Eq.~\eqref{ineq} implies that \f{\feps^*} does not satisfy Eq.~\eqref{weakcmpt} for all self-equilibrated fields~\f{\delta \tilde \fsigma}. In other words, only compatible strain fields satisfy Eq.~\eqref{weakcmpt}, which was to be demonstrated.\\
Equation~\eqref{weakcmpt} is classically known as Donati's theorem. It is closely related to St.~Venant's compatibility conditions and has been further developed and generalized ever since. For a complete and rigorous treatment and further reading, see \citet{ting1974st}, \citet{amrouche2006characterizations}, \citet{barbarosie2022donati}, and references therein.

\section{Two-scale simulation via Sequential Subspace Mode Adaptation}\label{app:twoscalealgo}
Algorithm \ref{algo:macroumat} shows how the Sequential Subspace Mode Adaptation is implemented with respect to each macroscopic integration point within a two-scale framework.

\begin{algorithm}[H]
\caption{Macroscopic integration point algorithm}
\label{algo:macroumat}
\footnotesize
\begin{algorithmic}

\State \textbf{Input:}{ $\bar\feps$, $\bar\feps_{n}$, $\Delta t$, $\boldsymbol{\xi}$, $\frac{\partial \boldsymbol{\xi}}{\partial \boldsymbol{\eta}}$, $\Nmd^0$ (number of base modes), $\Nmd^{\rm adptv}$ (number of adaptive modes), adapt\_modes}
\State \textbf{Output:} $\bar{\fsigma}, \barCa$ (macr.~algor.~tangent), $\boldsymbol{\xi}, \frac{\partial \boldsymbol{\xi}}{\partial \boldsymbol{\eta}}$
\State $\boldsymbol{\eta} \leftarrow \mathbf{0}$

\If{adapt\_modes}
    \State $n_\eta^0 \leftarrow \Nmd^0$
    \State Initialize adaptive reduced basis using the base modes: $\frac{\partial \boldsymbol{\xi}}{\partial \boldsymbol{\eta}}\leftarrow\begin{bmatrix}\mathbf I_{\Nmd^0}\\\mathbf 0\end{bmatrix}$
\Else
    \State $n_\eta^0 \leftarrow \Nmd^0+\Nmd^{\rm adptv}$
    \State $\frac{\partial \boldsymbol{\xi}}{\partial \boldsymbol{\eta}} \leftarrow \frac{\partial \boldsymbol{\xi}}{\partial \boldsymbol{\eta}}$
\EndIf

\For{$n_\eta = n_\eta^0, \dots, \Nmd^0+\Nmd^{\rm adptv}$}
    \State Construct reduced local strain fluctuation modes $\prescript{\eta}{}{\tcEps^{q}}$ from the local fluctuation basis $\tcEps^{q}$ generated offline:
    \State $\left(\prescript{\eta}{}{\tcEps^{q}_{1}},\dots,\prescript{\eta}{}{\tcEps^{q}_{n_\eta}}\right)\leftarrow\left(\prescript{}{}{\tcEps^{q}_{1}},\dots,\prescript{}{}{\tcEps^{q}_{N_{\rm md}}}\right)\frac{\partial \boldsymbol{\xi}}{\partial \boldsymbol{\eta}}$
    \State $\|\mathbf{R}\|_{\rm old} \leftarrow \infty;\quad n_{\rm iter} \leftarrow 0$

    \While{not converged}

        \State $\mathbf{R} \leftarrow \mathbf{0}$, 
               $\mathbf{K_{\rm red}} \leftarrow \mathbf{0}$

            \For{$q = 1, \dots, N_{\rm ip}$}

                \If{$n_{\rm iter} = 0$}
                    \State $\feps^{q} \leftarrow \bar{\feps}_{n} + \tcEps^{q} \boldsymbol{\xi}$
                \Else
                    \State $\feps^{q} \leftarrow \bar{\feps} + \tcEps^{q} \boldsymbol{\xi}$
                \EndIf

                \State $(\fsigma^{q}, {\Ca}^q) \leftarrow \text{evaluate material behavior based on } \feps^{q} \text{ and } \fX_n$

                \State $\mathbf{R} \leftarrow \mathbf{R} + \frac{\Omega_{q}}{\Omega} \T{\left(\tcEps^{q}\right)} \fsigma^{q}$

                \State $\mathbf{K}_{\rm red} \leftarrow \mathbf{K}_{\rm red} + \frac{\Omega_{q}}{\Omega} \T{\left(\prescript{\eta}{}{\tcEps^{q}}\right)} \left(\frac{\d\fsigma}{\d\feps}\right)^{q} \prescript{\eta}{}{\tcEps^{q}}$

            \EndFor

        \State $\mathbf{R}_{\rm red} \leftarrow \left(\frac{\partial \boldsymbol{\xi}}{\partial \boldsymbol{\eta}}\right)^T \mathbf{R}$

        \If{$\|\mathbf{R}_{\rm red}\| < \text{tol}$ and $n_{\rm iter} > 0$} 
            \State \textbf{exit}
        \EndIf

        \If{$n_{\rm iter} < 2$ or $\|\mathbf{R}_{\rm red}\| < \|\mathbf{R}\|_{\rm old}$}

            \State $\|\mathbf{R}\|_{\rm old} \leftarrow \|\mathbf{R}_{\rm red}\|$

            \State Solve: $\mathbf{K}_{\rm red} \Delta \boldsymbol{\eta} = -\mathbf{R}_{\rm red}$

            \State $\boldsymbol{\eta} \leftarrow \boldsymbol{\eta} + \Delta \boldsymbol{\eta};\quad\Delta \boldsymbol{\xi} \leftarrow \frac{\partial \boldsymbol{\xi}}{\partial \boldsymbol{\eta}} \Delta \boldsymbol{\eta};\quad\boldsymbol{\xi} \leftarrow \boldsymbol{\xi} + \Delta \boldsymbol{\xi}$

        \Else{ Reduce step size}

            \State $\boldsymbol{\xi} \leftarrow \boldsymbol{\xi} - \Delta \boldsymbol{\xi};\quad\Delta \boldsymbol{\xi} \leftarrow \frac{1}{2}\Delta \boldsymbol{\xi};\quad\boldsymbol{\xi} \leftarrow \boldsymbol{\xi} + \Delta \boldsymbol{\xi}$

        \EndIf

        \State $n_{\rm iter} \leftarrow n_{\rm iter} + 1$

    \EndWhile

    \If{$n_\eta < \Nmd^0+\Nmd^{\rm adptv}$}
        \State Expand: $\frac{\partial \fxi}{\partial \feta}\leftarrow \left(\frac{\partial \fxi}{\partial \feta},\fR\right)$
    \EndIf

\EndFor

\State $\bar{\fsigma} = \avg{\fsigma} = \frac{1}{\Omega} \sum_{q} \Omega_{q} \fsigma^{q}$

\State Macroscopic tangent:
$\barCa=\langle\Ca\rangle-\sum_{k,l}\left(K^{\rm red -1}\right)_{kl}\langle\Ca:\prescript{\eta}{}{\tcEps_k}\rangle\otimes\langle\Ca:\prescript{\eta}{}{\tcEps_l}\rangle$

\end{algorithmic}
\end{algorithm}

\section{Enhanced Levenberg Marquardt damping}\label{appdamping}
The Levenberg Marquardt damping described in \citet{wulfinghoff2026computational} (Eq.~(53)) is further extended by adding the following matrix to the approximate Hessian \f{\hat \fH} (besides the damping described in that work):
\begin{equation}
    \begin{pmatrix}
        \fD^\circ & \fzero & \dots & \fzero\\
        \fzero & \fD^\circ& \dots & \fzero\\
        \vdots & \vdots & \ddots & \vdots\\
        \fzero & \fzero & \dots & \fD^\circ
    \end{pmatrix}, \ \ \ 
    \fD^\circ = \begin{pmatrix}
        \lambda^\circ & \lambda^\circ & \lambda^\circ & 0 & 0 & 0\\
        \lambda^\circ & \lambda^\circ & \lambda^\circ & 0 & 0 & 0\\
        \lambda^\circ & \lambda^\circ & \lambda^\circ & 0 & 0 & 0\\
        0 & 0 & 0 & 0 & 0 & 0\\
        0 & 0 & 0 & 0 & 0 & 0\\
        0 & 0 & 0 & 0 & 0 & 0
    \end{pmatrix}.
\end{equation}
In this work, the additional volumetric damping parameter is set to~\f{\lambda^\circ=5\lambda}.
\section{Mean strain stabilized element formulation}\label{appmeanstrain}
The stabilized mean strain element is formulated based on the following weak form:
\begin{align}
    \int\limits_\Omega (\fsigma^*+\ffC_0:(\feps-\feps^*)) : \delta \feps \d \Omega=0,\\ 
    \int\limits_\Omega \delta \fsigma^* : (\feps-\feps^*) \d \Omega =0,\\
    \int\limits_\Omega \delta \feps^* : ( \fsigma(\feps^*,\fX_n) - \fsigma^* - \ffC_0:(\feps-\feps^*) ) \d \Omega=0
\end{align}
with~\f{\feps=\fsym{\grad{\fu}}}. The displacement field~\f{\fu} is discretized by conventional quadratic shape functions, while the stress~\f{\fsigma^*} and the assumed strain~\f{\feps^*} are approximated element wise constant and standard 4-point integration is used for all three aforementioned integrals on the element level. The reference stiffness is chosen as \f{\ffC_0=\lambda_0\fI\otimes\fI+ 2\mu_0\ffI}, where \f{\lambda_0= 2\mu_0=\sigma^{\rm c}/\varepsilon^{\rm c}} are chosen constant. Here, \f{\sigma^{\rm c}} and~\f{\varepsilon^{\rm c}} denote characteristic stress and strain values (in this work \f{\sigma^{\rm c}=\sigma_{\rm y0}} and \f{\varepsilon^{\rm c}=0.05}). For more advanced formulations and further reading on the topic, refer to \citet{nguyen2018modification}, \citet{wulfinghoff2018model} and references therein.

  \section{Cost function minimization and derivatives}
  \label{APPCostfunctionMinimizationDerivatives}
  The following sections will show the linearizations used for cost function minimization in combination with internal variables.

  \subsection{Derivatives of cost function $c$ with respect to $\tcfE$}
  \label{APPCostFunctionMinGeneral}
  The cost function $c(\tcfE)$ given in Eq.~\eqref{cstfctn}, has to be minimized with respect to the $\tcfE_{k}^{q}$. In general, the individual parts coming form the two contributions in Eq. \eqref{eq:HM_Contributions}, in the following denoted as $c^{R}$ and $c^{\bar\sigma}$, can be considered independently of each other, as
  \begin{equation}
    \Ds\pd{c(\tcfE)}{\tcfE_{k}^{q}}=\Ds\pd{c^{R}}{\tcfE_{k}^{q}}+\Ds\pd{c^{\bar\sigma}}{\tcfE_{k}^{q}}.
  \end{equation}
  The linearization of the first part $c^{R}$ is given by
  \begin{equation}
    \Ds\pd{c^{R}}{\tcfE_{k}^{q}}=\sum_{s=1}^{N_{\rm sim}}\sum_{n=0}^{N_{t}^{s}-1}\sum_{m=1}^{N_{\rm md}}\frac{1}{\Omega^{2}}R^{s}_{m(n+1)}\Ds\pd{R_{m(n+1)}^{s}}{\tcfE_{k}^{q}},
  \end{equation}
  with $\Ds\pd{R_{m(n+1)}^{s}}{\tcfE_{k}^{q}}$ as (see also \ref{APPLinearizationOfSigma})
  \begin{equation}\label{EQDerivativeResidualEpsModes}
    \Ds\pd{R_{m(n+1)}^{s}}{\tcfE_{k}^{q}}=\T{\biggl(\Ds\pd{\fsigma_{n+1}^{qs}}{\tcfE_{k}^{q}}\biggr)}:\tcfE_{m}^{q}\Omega^{q}+\fsigma_{n+1}^{qs}\delta_{km}\Omega^{q}.
  \end{equation}
  The linearization of the second part $c^{\bar\sigma}$ reads
  \begin{equation}
    \Ds\pd{c^{\bar\fsigma}}{\tcfE_{k}^{q}}=\sum_{s=1}^{N_{\rm sim}}\sum_{n=0}^{N_{t}^{s}-1}\T{\biggl(\Ds\pd{\Delta\bar\fsigma_{n+1}^{s}}{\tcfE_{k}^{q}}\biggr)}:\Delta\bar\fsigma_{n+1}^{s},
  \end{equation}
  with $\Delta\bar\fsigma_{n+1}^{s}=\langle\fsigma_{n+1}^{s}\rangle-\bar\fsigma_{n+1}^{s}$ and $\Ds\pd{\Delta\bar\fsigma_{n+1}^{s}}{\tcfE_{k}^{q}}$ as
  \begin{equation}\label{EQDerivativeDeltaSigmaEpsModes}
    \Ds\pd{\Delta\bar\fsigma_{n+1}^{s}}{\tcfE_{k}^{q}}=\frac{1}{\Omega}\sum_{p=1}^{N_{\rm ip}}\Ds\pd{\fsigma_{n+1}^{ps}}{\tcfE_{k}^{q}}\Omega^{p}=\frac{\Omega^{q}}{\Omega}\sum_{p=1}^{N_{\rm ip}}\Ds\pd{\fsigma_{n+1}^{ps}}{\tcfE_{k}^{q}}.
  \end{equation}
  In order to ensure that the constraint given in Eq.~\eqref{mdeavg} is valid, a reduced version of modes without the last component is given by $\utcE_{\rm red}$:
  \begin{equation}
    \utcE_{\rm red}=\begin{pmatrix}\utcE^{1}\\.\\.\\.\\\utcE^{N_{\rm ip}-1}\end{pmatrix}.
  \end{equation}
  The linearization of cost function $c$ results in the reduced form:
  \begin{align}
    \Delta c&=\Ds\pd{c(\utcE)}{\utcE}\Delta\utcE\notag\\
    &=\sum_{q=1}^{N_{\rm ip}-1}\Ds\pd{c}{\utcE^{q}}\Delta\utcE^{q}+\Ds\pd{c}{\utcE^{N_{\rm ip}}}\Delta\utcE^{N_{\rm ip}}\notag\\
    &=\sum_{q=1}^{N_{\rm ip}-1}\underbrace{\biggl(\Ds\pd{c}{\utcE^{q}}-\frac{\Omega^{q}}{\Omega^{N_{\rm ip}}}\Ds\pd{c}{\utcE^{N_{\rm ip}}}\biggr)}_{\biggl(\Ds\pd{c}{\utcE_{\rm red}}\biggr)^{q}}:\Delta\utcE^{q}
  \end{align}

  \subsection{Linearization of $\fsigma_{n+1}^{qs}$}
  \label{APPLinearizationOfSigma}
  The linearization of $\fsigma_{n+1}^{qs}$ is given by
  \begin{align}\label{EQLinearizationSigma}
    \Delta\fsigma_{n+1}^{qs}=&\Delta(\fsigma^{q}(\feps_{n+1}^{qs}{,}\fX_{n}^{qs}))\notag\\
    =&\underbrace{\Ds\pd{\fsigma^{q}(\feps_{n+1}^{qs}{,}\fX_{n}^{qs})}{\feps_{n+1}}}_{\rm algo.\:tangent}:\underbrace{\Delta\feps_{n+1}^{qs}}_
    {\sum_{k=1}^{N_{\rm md}}\xi_{k\,(n+1)}^{s}\Delta\blds{\tcfE}_{k}^{q}}
    +\Ds\pd{\fsigma^{q}(\feps_{n+1}^{qs}{,}\fX_{n}^{qs})}{\fX_{n}}\underbrace{\Delta\fX_{n}^{qs}}_
    {\sum_{k=1}^{N_{\rm md}}\frac{\partial\blds\fX_{n}^{qs}}{\partial\blds\tcfE_{k}^{q}}:\Delta\blds\tcfE_{k}^{q}}
    \notag\\
    =&\sum_{k=1}^{N_{\rm md}}\underbrace{\biggl(\xi_{k\,(n+1)}^{s}\Ds\pd{\fsigma^{q}(\feps_{n+1}^{qs}{,}\fX_{n}^{qs})}{\feps_{n+1}}+\Ds\pd{\fsigma^{q}(\feps_{n+1}^{qs}{,}\fX_{n}^{qs})}{\fX_{n}}\Ds\pd{\fX_{n}^{qs}}{\tcfE_{k}^{q}}\biggr)}_
    {\frac{\partial\blds\fsigma_{n+1}^{qs}}{\partial\blds\tcfE_{k}^{q}}}
    :\Delta\tcfE_{k}^{q}.
  \end{align}
  Further, $\fX_{n+1}^{qs}=\fX^{q}(\feps_{n+1}^{qs}{,}\fX_{n}^{qs})$ has to be linearized to complete the set of equations needed to perform the cost function minimization in Eq.~\eqref{cstfctn}:
  \begin{align}\label{EQLinearizationInternalVariableVector}
    \Delta\fX_{n+1}^{qs}=&\Delta(\fX^{q}(\feps_{n+1}^{qs}{,}\fX_{n}^{qs}))\notag\\
    =&\Ds\pd{\fX^{q}}{\feps_{n+1}}:\underbrace{\Delta\feps_{n+1}^{qs}}_
    {\sum_{k=1}^{N_{\rm md}}\xi_{k\,(n+1)}^{s}\Delta\blds\tcfE_{k}^{q}}
    +\Ds\pd{\fX^{q}}{\fX_{n}}\underbrace{\Delta\fX_{n}^{qs}}_
    {\sum_{k=1}^{N_{\rm md}}\frac{\partial\blds\fX_{n}^{qs}}{\partial\blds\tcfE_{k}^{q}}:\Delta\blds\tcfE_{k}^{q}}
    \notag\\
    =&\sum_{k=1}^{N_{\rm md}}\underbrace{\biggl(\Ds\pd{\fX^{q}(\feps_{n+1}^{qs}{,}\fX_{n}^{qs})}{\feps_{n+1}}\xi_{k\,(n+1)}^{s}+\Ds\pd{\fX^{q}(\feps_{n+1}^{qs}{,}\fX_{n}^{qs})}{\fX_{n}}\Ds\pd{\fX_{n}^{qs}}{\tcfE_{k}^{q}}\biggr)}_
    {\frac{\partial\blds\fX_{n+1}^{qs}}{\partial\blds\tcfE_{k}^{q}}}
    :\Delta\tcfE_{k}^{q}.
  \end{align}
  In this contribution the following non-standard partial derivatives occur with respect to the material laws introduced in Sec.~\ref{sectConst}:
  \begin{equation}
      \Ds\pd{\fsigma^{q}(\feps_{n+1}^{qs}{,}\fX_{n}^{qs})}{\fX_{n}}=\begin{pmatrix}\displaystyle\frac{\partial\fsigma^{q}}{\partial\fepsp_{n}} \\ \displaystyle\frac{\partial\fsigma^{q}}{\partial\alpha_{n}}\end{pmatrix}
  \end{equation}
  and
  \begin{equation}
      \Ds\pd{\fX^{q}(\feps_{n+1}^{qs}{,}\fX_{n}^{qs})}{\fX_{n}}=\begin{pmatrix}
        \displaystyle\frac{\partial\fepsp}{\partial\fepsp_{n}} & \displaystyle\frac{\partial\fepsp}{\partial\alpha_{n}}\\
        \displaystyle\frac{\partial\alpha}{\partial\fepsp_{n}} & \displaystyle\frac{\partial\alpha}{\partial\alpha_{n}}
      \end{pmatrix}.
  \end{equation}
  The individual partial derivatives are shown here. For simplicity only the first one is written with all details, introducing short notations used in the following. At first the partial derivatives of \f{\fsigma^{q}} with respect to the previous state, given by \f{\fepsp_{n}} and \f{\alpha_n} are shown:
  \begin{align}
      \frac{\partial\fsigma^{q}}{\partial\fepsp_{n}}=-\frac{\partial\fsigma^{q}}{\partial\feps}=
      &\underbrace{-\lambda\fI\otimes\fI-2\mu\ffI^{\rm s}}_{-\ffC}
      -6\mu^{2}\underbrace{\left[\frac{\overbrace{\Delta t \dot\varepsilon_{0}\frac{p}{\sigma^{D}}\left\langle\frac{\sqrt{\frac{3}{2}}\|(\fsigma')^{q}\|-\sigma_{\rm y}}{\sigma^{D}}\right\rangle^{p-1}}^{b}}{-\frac{3}{2}b\biggl(2\mu+\frac{2}{3}\sigma_{\rm y}^{'}(\alpha)\biggr)-1}\right]}_{w}\fN\otimes\fN \notag\\ 
      &+\frac{4\mu^{2}\Delta\gamma}{\|{(\fsigma')^{\mathrm{Tr},q}}\|}\biggl( \ffP^{\rm s}-\fN\otimes\fN\biggr),
  \end{align}
  \begin{equation}
      \frac{\partial\fsigma^{q}}{\partial\alpha_{n}}=-2\sqrt{\frac{3}{2}}\mu w\sigma_{\rm y}^{'}(\alpha)\fN.
  \end{equation}
  Above and in the following \f{\ffI^{\rm s}} denotes the symmetric fourth-order identity tensor, \f{\ffP^{\rm s}} the deviatoric projection operator and \f{\fN} the direction of flow. Next, the partial derivatives of the internal variables \f{\fX_n} are given:
  \begin{equation}
      \frac{\partial\fepsp}{\partial\fepsp_{n}}=-\frac{\partial\fepsp}{\partial\feps}=3\mu w\fN\otimes\fN+\frac{2\mu}{\|{(\fsigma')^{\mathrm{Tr},q}}\|}\biggl(\ffP^{\rm s}-\fN\otimes\fN\biggr)+\ffI^{\rm s},
  \end{equation}
  \begin{equation}
      \frac{\partial\fepsp}{\partial\alpha_{n}}=\sqrt{\frac{3}{2}}w\sigma_{\rm y}^{'}(\alpha)\fN,
  \end{equation}
\begin{equation}
      \frac{\partial\alpha}{\partial\fepsp_{n}}=-\frac{\partial\alpha}{\partial\feps}=3\sqrt{\frac{2}{3}}\mu w\fN,
  \end{equation}
  \begin{equation}
      \frac{\partial\alpha}{\partial\alpha_{n}}=w\sigma_{\rm y}^{'}(\alpha)+1.
  \end{equation}
    
\end{appendices}
\bibliographystyle{elsarticle-harv}
\bibliography{lit_zotero}


\end{document}